\documentclass[twocolumn]{aastex631}

\usepackage{rotating}

\accepted{March 3, 2025}

\submitjournal{ApJ}

\shorttitle{The dwarf irregular galaxy NGC~6822}
\shortauthors{Tantalo et al.}

\graphicspath{{./}{figures/}}

\begin{document}

\title{The dwarf irregular galaxy NGC~6822. II. Young, intermediate and old stellar populations: comparison between theory and observations}

\correspondingauthor{Maria Tantalo}
\email{maria.tantalo@inaf.it}

\author[0000-0002-6829-6704]{Maria Tantalo}
\affiliation{INAF - Osservatorio Astronomico di Roma, via Frascati 33, I-00078 Monte Porzio Catone, Italy}
\affiliation{Instituto de Astrofísica de Canarias, Calle Via Lactea s/n, E-38205 La Laguna, Tenerife, Spain}
\affiliation{Departamento de Astrofísica, Universidad de La Laguna (ULL), E-38200, La Laguna, Tenerife, Spain}

\author{Giuseppe Bono}
\affiliation{Dipartimento di Fisica, Università di Roma Tor Vergata, via della Ricerca Scientifica 1, I-00133 Roma, Italy}
\affiliation{INAF - Osservatorio Astronomico di Roma, via Frascati 33, I-00078 Monte Porzio Catone, Italy}

\author{Maurizio Salaris}
\affiliation{Astrophysics Research Institute, Liverpool John Moores University, 146 Brownlow Hill, Liverpool L3 5RF, United Kingdom}
\affiliation{INAF - Osservatorio Astronomico d’Abruzzo, via M. Maggini, s/n, I-64100, Teramo, Italy}

\author{Adriano Pietrinferni}
\affiliation{INAF - Osservatorio Astronomico d’Abruzzo, via M. Maggini, s/n, I-64100, Teramo, Italy}

\author{Matteo Monelli}
\affiliation{INAF - Osservatorio Astronomico di Roma, via Frascati 33, I-00078 Monte Porzio Catone, Italy}
\affiliation{Instituto de Astrofísica de Canarias, Calle Via Lactea s/n, E-38205 La Laguna, Tenerife, Spain}
\affiliation{Departamento de Astrofísica, Universidad de La Laguna (ULL), E-38200, La Laguna, Tenerife, Spain}

\author{Michele Fabrizio}
\affiliation{Space Science Data Center, via del Politecnico snc, I-00133 Roma, Italy}
\affiliation{INAF - Osservatorio Astronomico di Roma, via Frascati 33, I-00078 Monte Porzio Catone, Italy}

\author{Vittorio F. Braga}
\affiliation{INAF - Osservatorio Astronomico di Roma, via Frascati 33, I-00078 Monte Porzio Catone, Italy}

\author{Annalisa Calamida}
\affiliation{Space Telescope Science Institute, 3700 San Martin Drive, Baltimore, MD 21218}

\author{Massimo Dall'Ora}
\affiliation{INAF - Osservatorio Astronomico di Capodimonte, Salita Moiariello 16, I-80131 Napoli, Italy}

\author{Valentina D'Orazi}
\affiliation{Dipartimento di Fisica, Università di Roma Tor Vergata, via della Ricerca Scientifica 1, I-00133 Roma, Italy}
\affiliation{INAF - Osservatorio Astronomico di Padova, vicolo dell'Osservatorio 5, I-35122 Padova, Italy}

\author{Ivan Ferraro}
\affiliation{INAF - Osservatorio Astronomico di Roma, via Frascati 33, I-00078 Monte Porzio Catone, Italy}

\author{Giuliana Fiorentino}
\affiliation{INAF - Osservatorio Astronomico di Roma, via Frascati 33, I-00078 Monte Porzio Catone, Italy}

\author{Giacinto Iannicola}
\affiliation{INAF - Osservatorio Astronomico di Roma, via Frascati 33, I-00078 Monte Porzio Catone, Italy}

\author{Massimo Marengo}
\affiliation{Department of Physics, Florida State University, 77 Chieftain Way, Tallahassee, FL32306, USA}

\author{Noriyuki Matsunaga}
\affiliation{Department of Astronomy, School of Science, The University of Tokyo, 7-3-1 Hongo, Bunkyo-ku, Tokyo 113-0033, Japan}

\author{Joseph P. Mullen}
\affiliation{Department of Physics and Astronomy, Vanderbilt University, Nashville, TN 37240, USA}

\author{Peter B. Stetson}
\affiliation{Herzberg Astronomy and Astrophysics, National Research Council, 5071 West Saanich Road, Victoria, British Columbia V9E 2E7, Canada}

\begin{abstract}

This paper presents a quantitative analysis of the stellar content in the Local Group dwarf irregular
galaxy NGC~6822 by comparing stellar evolution models and observations in color-magnitude diagrams 
(CMDs) and color-color diagrams (CC-Ds). Our analysis is based on optical ground-based $g,r,i$ photometry, and 
deep archive HST photometry of two fields in the galaxy disk.
We compared young, intermediate-age, and old stellar populations with isochrones from the
BaSTI-IAC library and found that NGC~6822 hosts a quite metal-rich ([Fe/H] = $-$0.7 to $-$0.4)
young component with an age ranging from 20 to 100 Myr. The intermediate-age population experienced a
modest chemical enrichment between 4 and 8~Gyr ago while stars older than 11 Gyr have a low metal
abundance ([Fe/H]$\sim-$1.70).
We also identified the AGB clump population with a luminosity peak at $i\sim 23.35$ mag.
Our analysis of both the CMD and the optical-NIR-MIR CC-Ds of AGB oxygen- and
carbon-rich stars, using the PARSEC+COLIBRI isochrones with and without circumstellar dust, reveal that this stellar component exhibits a spread in age from 1 to 2 Gyr and in metallicity between [Fe/H]=$-$1.30 and $-$1.70.
The stellar models we used reproduce very well the two distinct color sequences defined by AGB O- and C-rich
stars in the various optical-NIR-MIR CC-Ds, suggesting that they are reliable diagnostics to identify and characterise intermediate-age stellar populations.
However, we also find that evolutionary prescriptions in the optical $i$-($r-i$) CMDs predict, at fixed color, systematically lower luminosities than observed AGB stars.

\end{abstract}


\defcitealias{Tantalo22}{Paper I}
\defcitealias{Letarte02}{L02}

\section{Introduction} \label{sec:intro}

The Local Group (LG) hosts six dwarf irregular (dIrr) galaxies that are fundamental laboratories of low-intensity and/or low-metallicity star formation in stellar systems 
of the local Universe: Magellanic Clouds (MCs), WLM, IC~10, IC~1613 and NGC~6822. The MCs are the most prominent and massive dIrr, but current theoretical and observational evidence indicates that they are in a mutual interaction with each other and probably tidally perturbed by the Milky Way \citep[MW; see][and references therein]{Besla16,Zivick18,Choi22}.
The others are more isolated and considered unique analogs of the star forming galaxies populating the universe at the peak of cosmic star formation history \citep{Stott13, Du20}.
Among these, NGC~6822 is the stellar system most closely resembling the MCs in mass and luminosity.
Indeed, there is evidence of a sizable sample of star clusters \citep{Hwang11,Huxor13}, 
and of a disk with an ongoing star formation activity 
\citep[see i.e. \citeauthor{Tantalo22} \citeyear{Tantalo22}, hereinafter~\citetalias{Tantalo22};][]{Lenkic23}. Moreover, NGC~6822 is the closest dIrr to the MW after the MCs \citep[d $\sim$ 510 $\pm$ 10 kpc;][]{Fusco14}, and one of the most investigated dIrr galaxies, due its position in the sky (R.A.(J2000) = $19^{h} 44^{m} 58.56^{s}$, 
DEC(J2000) = $-14^{d} 47^{m} 34.8^{s}$; \citetalias{Tantalo22}) 
that makes it observable from both the northern and the southern hemispheres. 
Although it is commonly considered isolated, more recent studies on its orbital history speculated that it could have crossed the virial radius of the MW $\sim$3 Gyr ago \citep*{Zhang21,Bennet23}.

Owing to its low Galactic latitude \citep[$l=25.4^{\degr}$ and $b=-18.4^{\degr}$,][]{Mateo98} 
NGC~6822 is affected by a fair amount of extinction. 
\citet*{Schlegel98}'s map provides a mean  
reddening along the line of sight  $E(B-V)=0.24$ mag. Several studies have shown that the 
galaxy is also affected by differential reddening, varying from $E(B-V)=0.37$ mag near 
the center to $E(B-V)=0.30$ mag in the outermost regions \citep*{Massey95, Gallart96a, Gieren06, Fusco12}.
The true distance modulus has been measured by using several different methods. The most 
recent estimates come from the tip of the red giant branch 
\citep[TRGB; $\mu$ = 23.54 $\pm$ 0.05 mag,][]{Fusco12}, from the period-luminosity relation of 
classical Cepheids \citep[$\mu$ = 23.38 $\pm$ 0.04 mag,][]{Rich14} and from the Mira variables 
\citep[$\mu$ = 23.38 $\pm$ 0.16 mag,][]{White13}.

The stellar content of NGC~6822 has been studied in detail by several authors during the last century. For example, 
Hubble succeeded in the identification of Classical Cepheids soon after his seminal investigation on M31 \citep{Hubble25}. 
The emerging view is that NGC~6822 
hosts stellar populations ranging from very young (a few Myr) to intermediate-age (a few Gyr) and old (t$>$10 Gyr) low-mass stars (\citetalias[][and references therein]{Tantalo22}), suggesting that NGC~6822 experienced multiple star formation episodes over its entire life. 
A detailed star formation history (SFH) of the galaxy was determined by \citet{Gallart96b,Gallart96c}, 
who found that the galaxy has most likely started to form stars from a low metallicity gas ($Z_i$ = 0.0001-0.0004) around 12-15 Gyr ago. They also found that the star formation rate (SFR) began 
to grow in last 400 Myr and there has been a further enhancement in the last 100-200 Myr over the whole body of the galaxy. 
Several years later, \citet{Wyder03} derived the SFHs of four fields with optical images collected with HST/WFPC2 covering the minor axis of NGC~6822. They found that the SFHs in the investigated fields are similar for ages older than $>1$ Gyr. The SFRs are fairly constant up to  $t\sim600$ Myr, but in the most recent epochs they show a drop of a factor of $\sim 2-4$ in the more external fields and a well defined increase in the field centered on the bar.
Moreover, they also compared the derived SFRs with those expected from the typical gas surface densities at the galactocentric radii of the galaxy and they found that they are comparable, thus suggesting that no large-scale redistribution either of gas or of stars is required to take account for the inferred SFRs.
More recently, \citet{Cannon12} studied images of six HST/ACS fields along the HI major axis 
of NGC~6822 and derived their SFHs, supporting the results of \citet{Gallart96b,Gallart96c}. 
They argued that a high fraction of the stars formed between 14 and 6 Gyr ago and half of the total stellar mass was built up in the last $\sim$5 Gyr. Furthermore, they also suggested that the SFR increased in the last $\sim$50 Myr.
Besides, the outer ACS pointings showed more old stars than in the inner ones, suggesting an active star-forming region shrinking inward with time.
\citet{Fusco14} used the same data set provided by \citet{Cannon12}, but using a different 
technique, to compute the SFHs of the six fields. The solutions are consistent with 
\citet{Cannon12}, and revealed that the star formation activity in the selected fields, but one, 
has slowly decreased in the last 1~Gyr. Also, they derived the age-metallicity 
relation (AMR) for the last 5 Gyr disclosing that the metallicity raised with time from 
[Fe/H] $\simeq -$1 to [Fe/H] $\simeq -$0.5 in all the six fields.
The main limitation of these detailed investigations is that they only 
cover a minor portion of the galaxy, with small fields typically located across the 
central regions.

Young stellar populations in NGC~6822 have been widely examined 
in the near-infrared (NIR), mid-infrared (MIR) and far-infrared (FIR) bands. 
A huge number of Young Stellar Objects (YSOs) have been identified and characterized by using both IR observations 
\citep*[see i.e.][]{Jones19,Hirsc20,Lenkic23} 
and machine learning technique \citep{Kinson21}, 
a clear indication that the galaxy is still actively forming stars.

Intermediate-age stellar tracers, such as red clump (RC) and asymptotic giant branch (AGB) stars, have also been extensively studied in NGC~6822. AGB stars are popular stellar tracers that can be easily identified in stellar systems hosting stellar populations older than $\sim$0.5 Gyr. However, in complex stellar systems that have experienced recent star formation events, like NGC~6822, the AGB component is typically dominated by intermediate-mass stars (2 $\le$ M/M$_{\odot} \le$ 8). Note that AGB stars are among the most important contributors to the IR integrated light of a galaxy \citep{Renzini86} and play a key role in the chemical enrichment of galaxies, since they produce neutron-capture elements.
The thermally pulsing AGB (TP-AGB) stars are classified according to their surface carbon-to-oxygen abundance ratios (C/O) in three different groups: C-type (C/O$>$1), O-type (C/O$<$1, denoted  also as M-type) and S-type (C/O$\sim$1). Their surface chemical composition is affected by complex physical mechanisms (e.g. 3rd dredge-up, convective overshooting, mass loss, hot bottom burning) mainly depending on the stellar mass and the metallicity of the environment in which they formed. The population ratio between C- and M-type (C/M) stars can be adopted to trace the metal content of the intermediate-age stellar populations \citep{Battinelli05,Cioni09} and metal-poor/metal-intermediate stellar systems, like NGC~6822, are fundamental laboratories to trace the variation of the population ratio \citep[see i.e.][]{Weiss09}.
Furthermore, photometric investigations on their spatial distributions indicated that the C-rich stars are more 
centrally concentrated than the O-rich stars (\citetalias[][and references therein]{Tantalo22}). This means 
that a comprehensive analysis of AGB stars covering the entire body of NGC~6822 can provide solid constraints 
on the role played by the environment on AGB evolution \citep{Letarte02,Kang06,Kacharov12,Sibbons12,Sibbons15,Hirsc20,Nally24}.

\citet{Nally24} recently analyzed NIR and MIR images from the James Webb Space Telescope (JWST) to investigate AGB stars in NGC~6822, particularly to identify the AGB clump. However, their analysis of AGB stars was mainly 
based on a variety of NIR/MIR  Color-Magnitude Diagrams (CMDs) and luminosity functions. 
Furthermore, a complete identification of these evolved stars can lead to a better understanding 
of the dust production process in a galaxy. There is mounting observational evidence that 
AGB stars are the primary sources of dust production in stellar systems 
\citep[i.e.][]{Boyer11,Boyer12,Boyer15a,Boyer15b,White18,Jones18,Nally24}.
Indeed, \citet{Boyer12} found that C-rich stars contribute $87\%-89\%$ of the total 
dust injection in the SMC against the $4\%$ of red super-giants (RSGs).

This is the second paper of a series mainly focused on the stellar content of NGC~6822. 
The photometric catalog was introduced in \citetalias{Tantalo22} together with solid 
photometric diagnostics for the identification of O- and C-rich AGB stars. 
Moreover, we used the population ratio C/M to constrain the mean metallicity of the intermediate-age stellar population.
Previous investigations of stellar populations in NGC~6822 based on HST data \citep[see i.e.][]{Cannon12,Fusco14},
were hampered by the limited area covered by the six ACS fields along the disk major axis, and by the limiting magnitude 
marginally reaching the oldest main sequence turn-off (MSTO) stars.

In the current investigation we take advantage of our optical ground-based data set, 
sampling nearly all of the galaxy, together with two very deep 
and accurate optical pointings based on ACS and WFC3 images 
to perform a more quantitative comparison in ground- and in space-based CMDs 
between cluster isochrones and observations. 
This comparison aims to provide constraints on 
the range in age and metallicity covered by old, intermediate, and young 
stellar populations. We also performed a detailed comparison in 
CMDs and in color-color diagrams (CC-Ds) with cluster isochrones, both 
taking into account and neglecting the presence of circumstellar dust in 
TP-AGB stars. Such comparisons between theory and 
observations are crucial to validate the photometric criteria adopted to 
select the candidate AGB stars.

The structure of the paper is as follows. Section~\ref{sec:pop} describes  
the selection criteria adopted to identify star samples, in the optical 
$gri$ catalog, representative of the young, intermediate and old stellar 
populations. It also shows the outcomes of both these sub-samples and 
HST CMDs analysis. Section~\ref{sec:agb} deals with AGB stars and the comparison with 
stellar evolution models either including or neglecting the presence of the dust.
In Section~\ref{sec:conclusion} we summarise the results and outline the near 
future plans of the current project.

\section{Observations} \label{sec:pop}

\begin{figure}[ht!]
  \centering
  \includegraphics[width=8.6cm]{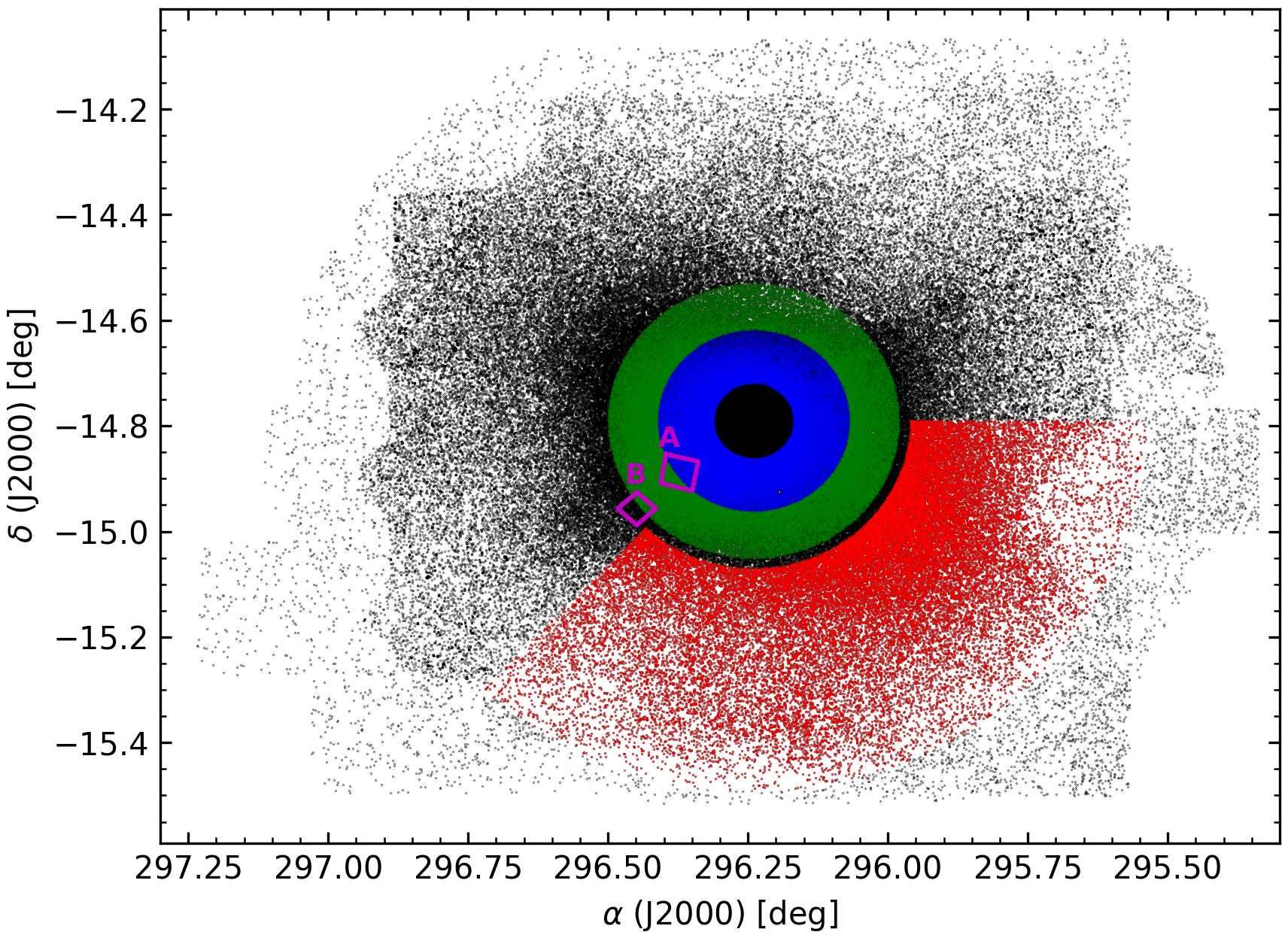}
  \includegraphics[width=8.7cm]{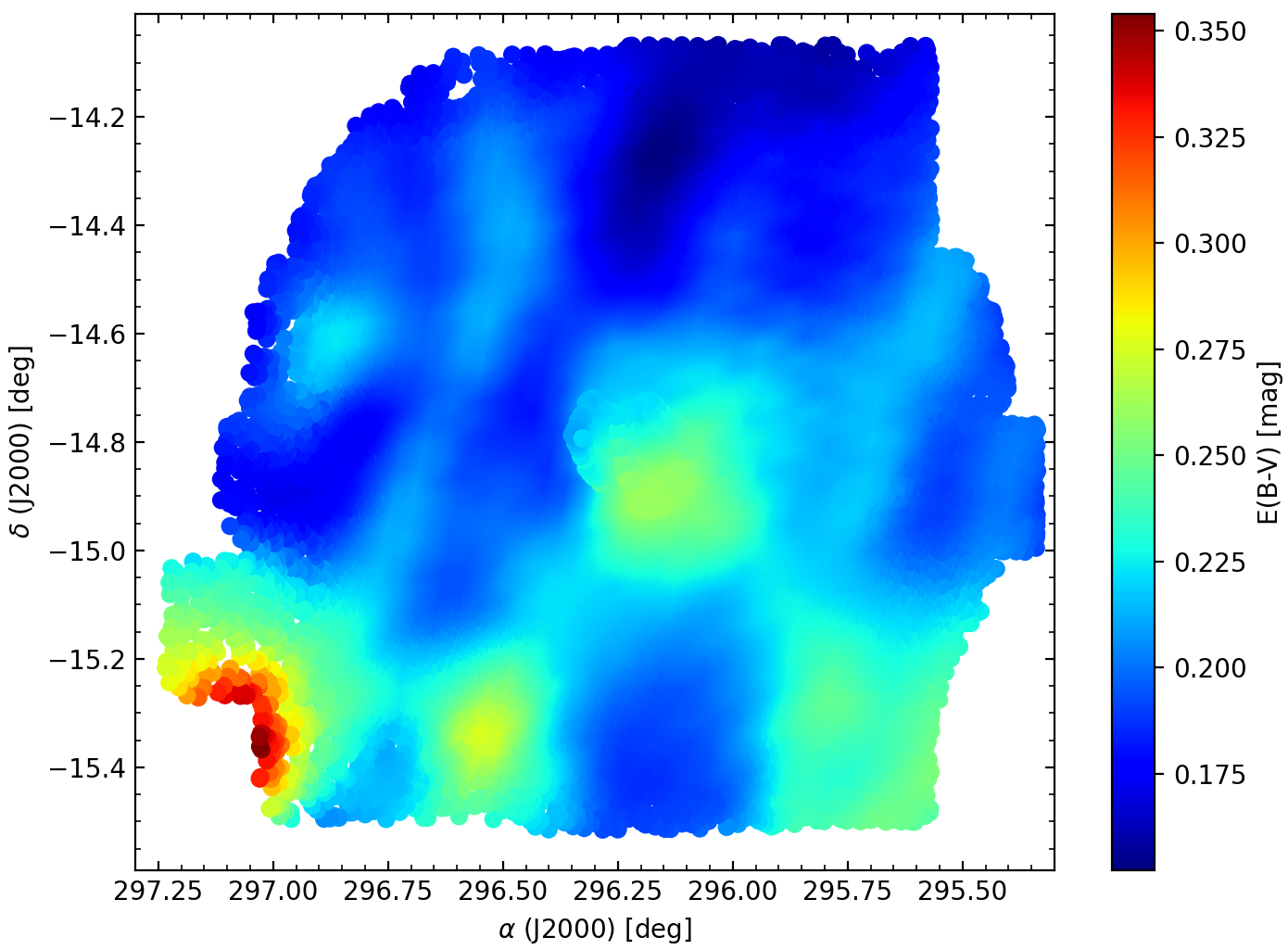} 
  \caption{Top: Sky coverage of the candidate galaxy stars (black dots) and the selection of young (blue dots), 
           intermediate-age (green dots) and old (red dots) samples (see text for more details) 
           to derive the $r$-($g-r$) CMDs in Figures~\ref{fig:CMDyoung}, ~\ref{fig:interm} and ~\ref{fig:CMDold}. 
           The magenta squares outline the boundaries of the HST fields A and B, as labeled.
  Bottom: Foreground reddening map of the sky region covered by the current dataset of NGC~6822.
           The map is color-coded according to the Galactic $E(B-V)$ values from \citet{Schlegel98} at the sky coordinates of the catalog.
           \label{fig:map}}
\end{figure}

The main sample of optical $g,r,i$ images includes mosaics collected with the Hyper-Suprime-Cam (HSC) at Maunakea Subaru Telescope.
This data were complemented by multi-band images acquired with the wide field imager 
MegaPrime available at the Canada-France-Hawaii Telescope (CFHT), 
with the Dark Energy Camera (DECam) on the Blanco telescope at Cerro Tololo 
Inter-American Observatory (CTIO), and with the Wide Field Camera (WFC) mounted on 
the Isaac Newton Telescope (INT) at La Palma. The entire data set covers a 2~deg$^2$ 
Field of View (FoV) across the center of the galaxy.
The data reduction was mainly performed with the DAOPHOT-ALLSTAR-ALLFRAME 
packages \citep{Stetson87, Stetson94}. 
The reader interested in a more detailed discussion concerning the software we adopted 
to perform PSF photometry together with the precision and the uniformity of the 
photometric zero-points across the FoV is referred to Appendix~\ref{appendix}.
To characterise the O- and C-rich AGB stars, the optical catalog was 
cross-correlated with NIR ($JHK$) photometry \citep{Sibbons12}, based on images 
collected with the Wide Field CAMera at the United Kingdom Infrared Telescope (UKIRT), 
and the MIR (3.6, 4.5, 5.8, 8.0 and 24~$\mu m$) photometry \citep{Khan15} collected with the Infrared Array 
Camera (IRAC) on board the Spitzer Space Telescope. The reader interested in more details 
about the optical-NIR-MIR catalogs is referred to Section~2 of \citetalias{Tantalo22}.
To overcome difficulties in the identification of different evolutionary features in the CMDs we are only using the catalog including candidate galaxy stars (see Section~3 in \citetalias{Tantalo22}).
The sky coverage from that catalog is presented in the top panel of Figure~\ref{fig:map} (black dots). 

We complemented ground-based photometry with deep archive HST data of an ACS and a WFC3 parallel field in the disk of NGC~6822, covering two pointings (Proposal ID=14191, PI A. Cole)\footnote{The archive HST data used in this paper can be found in MAST: \dataset[https://doi.org/10.17909/6vkj-nm47]{https://doi.org/10.17909/6vkj-nm47}.}, 
that overlap the areas Grid5 and Grid6 of the study performed by \citet{Cannon12}.
Data reduction followed the standard techniques based on the DAOPHOT/ALLFRAME code \citep{Stetson87,Stetson94}, using the same approach as in \citet{Monelli10b}. The single CTE-corrected FLC images were reduced individually, paying special care in deriving well-suited stars to model the Point-Spread Function (PSF). In a second step, images of the same WFC3 and ACS chip were combined to generate a global list of sources. 
The latter was fed to Allframe which improved the photometry through a simultaneous PSF-fitting on the objects of all the available frames for each detector. The photometric calibration was performed on the Vegamag system by utilizing the latest zero-points and the encircled energy corrections provided for the WFC3 and the ACS detectors\footnote{They are provided at \url{https://www.stsci.edu/hst/instrumentation/wfc3/data-analysis/photometric-calibration/} and \url{https://www.stsci.edu/hst/instrumentation/acs/data-analysis/zeropoints/}}.
The sky coverage of the two fields is displayed in Figure~\ref{fig:map} as magenta rectangles.
For the sake of clarity, hereinafter we will refer to the field closer to the galaxy center as field A, 
and to the field farther from the center as field B, as labeled in figure.
Note that the current HST data set is published here for the first time, and the derived CMD 
is more than $\sim$1 mag deeper than previous CMDs for NGC~6822 \citep[$F814W \sim$ 27 mag,][]{Fusco14} based on 
HST images. Indeed, the CMD we obtained approaches a limiting magnitude $F814W \sim 28$ mag, that is 
$\sim$1 mag fainter than the oldest MSTO of the galaxy as suggested by the theoretical isochrones 
(see Figs.~\ref{fig:HSTint} and \ref{fig:HSTold}).

To carry out a quantitative analysis of the stellar populations hosted by NGC~6822 
we took advantage of the sets of stellar isochrones available in the BaSTI-IAC database and the 
CMD 3.7 web interface\footnote{The interested reader is referred to 
\url{http://basti-iac.oa-abruzzo.inaf.it/index.html} and \url{http://stev.oapd.inaf.it/cgi-bin/cmd}}.
To perform a detailed comparison between theory and observations, we selected three different star samples that are representative of the young, intermediate-age and old stellar 
populations and they will be discussed in more detail in the 
subsections~\ref{sec:young}, \ref{sec:intermediate} and \ref{sec:old}. 

The top panel of Figure~\ref{fig:map} shows the sky coverage of the catalog 
of candidate galaxy stars (black dots) together with the young (a few Myr; blue), 
intermediate-age (a few Gyr; green), and old (t$>$10 Gyr; red) stellar samples.
The selection criteria for the young, intermediate and old stellar populations 
are based on a heuristic approach. Dating back to \citet{Mateo98,Stetson98,Monelli03} 
there is solid observational evidence that stellar populations in dwarf galaxies with complex 
star formation histories show well defined radial gradients. Young and intermediate 
stellar populations are more centrally concentrated and display a more flattened distribution, 
while old stellar populations have a more spherical distribution and reach the outskirt of 
the galaxy. Our findings on the spatial distributions and the structural parameters of different stellar tracers discussed in \citetalias{Tantalo22} (see Figure~16 of \citetalias{Tantalo22}) fully support this circumstantial evidence.
This is the main reason why we chose three different regions at increasing 
radial distance. The young stars have been selected in a region closer to the galaxy center.
Note that the more central galactic area was not included, since the photometry is 
less accurate due to crowding problems. To get a more homogeneous and accurate optical CMD 
we decided to use an annulus (blue area in the top panel of Figure~\ref{fig:map}) enclosed in an annulus with radius ranging from 0.07 and 0.17 degrees from the galaxy center
($\alpha$=296.24, $\delta$=-14.79; \citetalias{Tantalo22}). The young sample includes 211,896 stars.
The intermediate-age stellar tracers are more spatially extended, and also the RC stars 
steadily decrease in number towards the center due to the crowding and internal reddening
(see discussion in Section~5.3 of \citetalias{Tantalo22}). Therefore, they were selected in an annulus contiguous to the young stars and within a circle of 0.26 degrees in radius (green area in the top panel of Figure~\ref{fig:map}). We ended up with an intermediate-age sample of 86,809 stars.
For the old stellar populations we selected a portion of the galaxy 
with low reddening, since the limiting magnitude is fainter and the evolved sequence is 
characterized by narrower color distributions. The bottom panel of Figure~\ref{fig:map} shows 
the foreground reddening map of the sky region covered by the current dataset, color-coded
according to the Galactic $E(B-V)$ values from \citet{Schlegel98} at the sky coordinates of
the catalog. Note that in the innermost galaxy area the reddening is higher than in the
outermost part. We then selected an external annulus with a radius ranging from 0.28 to 0.70 degrees. We would also like to mention here that NGC~6822 hosts eight
globular clusters \citep[GCs;][]{Hubble25, Hwang11, Huxor13}, seven of which are located in the
northern regions. To avoid contamination from both the stellar populations of the GCs and
the disk of gas and young stars, we selected the old stars from a section of the external annulus
in the southwestern direction of the galaxy (red area in the top panel of Figure~\ref{fig:map}).
The sample contains 41,252 stars.
Moreover, the boundaries of the selected regions were adjusted to 
overlap with HST fields. Indeed, the HST field A is at the boundary between the young and 
the intermediate region, while the HST field B is at the boundary between the intermediate 
and the old region.

\section{Comparison with cluster isochrones } \label{sec:models}

In the following subsections we will compare the observed 
$r$-($g-r$) CMD of the young, intermediate and old stellar populations with 
cluster isochrones, and we will discuss in more detail their evolutionary features.

\subsection{Young stellar population} \label{sec:young}

\begin{figure}[ht!]
  \centering
  \includegraphics[width=8.7cm]{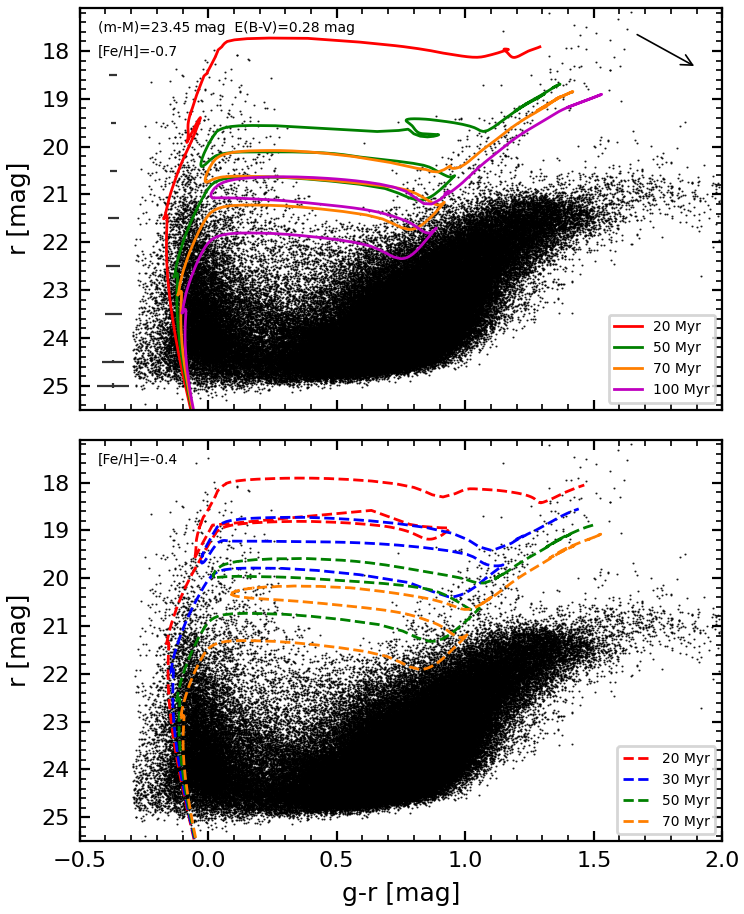}
  \caption{Top: comparison between the observed $r$-($g-r$) CMD of the young stars sample 
           (see Figure~\ref{fig:map}) and five scaled-solar BaSTI-IAC isochrones 
           for a fixed metallicity ([Fe/H]=$-0.7$) and different assumptions about age: 
           $20, 50, 70$ and $100$ Myr (see labels).
           The isochrones have been corrected for a true distance modulus of $23.45$ mag 
           and a reddening value of $E(B-V)=0.28$ mag.
           The error bars plotted on the left side of the CMD display the intrinsic errors both in magnitude and in color (summed in quadrature).
           The black arrow plotted in the upper right corner shows the reddening vector 
           for the extinction values adopted in the current analysis.
           Bottom: same as the top, but the isochrones are for a fixed metallicity of [Fe/H]=$-0.4$ and ages of $20, 30, 50$ and $70$ Myr. 
           \label{fig:CMDyoung}}
\end{figure}

Figure~\ref{fig:CMDyoung} shows the comparison between the observed $r$-($g-r$) CMD of the young stellar populations (see Figure~\ref{fig:map}) and stellar isochrones. 
To improve the comparison, the sources plotted in 
the CMD were selected according to their 
$\chi$ ($\chi < 2$) and 
$Sharpness $ ($-0.5 < Sharpness < 0.5$) values. 
These criteria allowed us to retain only 
stars with accurate photometry and neglect spurious sources and/or (partially) blended sources (this resulted in a subset of 160,520 sources from the original catalog).
We adopted scaled-solar isochrones from the BaSTI-IAC library \citep{Hidalgo18} 
and applied a true distance modulus of $\mu=23.45$ mag, and 
a mean $E(B-V)=0.28$ mag. 
The adopted distance modulus is the mean value 
between those obtained by \citet{Fusco12}, using the TRGB, 
and by \citet{Rich14}, using classical Cepheids. 
The reddening was estimated as the mean of the reddening values, retrieved from the map provided by \citet{Schlegel98}, across the galaxy region including the young stars (blue annulus in Figure~\ref{fig:map}).
The extinction in the individual photometric bands was calculated by using the \citet*{Cardelli89}'s reddening law ($A_g$=$1.18 A_V$, $A_r$=$0.88 A_V$). 
The metallicity values of the young isochrones have been chosen according to the spectroscopic measurements of a few supergiants available in literature, i.e. [Fe/H] = -0.49 $\pm$ 0.22 by \citet{Venn01} and [Z] = -0.52 $\pm$ 0.21 by \citet{Patrick15}.

The top panel of Figure~\ref{fig:CMDyoung} shows the comparison between four different stellar isochrones  
at a fixed metal content ([Fe/H]=-0.7) and the observed CMD. The youngest isochrone 
for 20 Myrs (red line) traces very well the blue edge of the young main sequence (YMS), while 
the older ones (t$\ge$50 Myrs) accounts for the helium (He) burning region, the so-called blue loop of young stars. 
Note that, as expected, the CMD of the young sample is contaminated by intermediate-age and old stars (i.e. RGB and AGB). Since they are distributed over the entire body of the galaxy, reducing their presence is quite challenging. Notwithstanding the foregoing, the young evolutionary tracers can be easily distinguished, because the red giant phases of these young
stellar structures are, as expected, systematically bluer than the red giant phases of 
both intermediate and old stellar populations. 
The agreement between theory and observations is quite good over the entire magnitude 
and color range of the quoted evolutionary phases. Indeed, more than 70\% of the red giants 
with $g$-$r$ colors between 1.0 and 1.5 mag and $r$ magnitudes between 18.5 and 20.5 mag are 
bracketed by isochrones with ages between 50 and 100 Myrs.

The bottom panel of Figure~\ref{fig:CMDyoung} shows the comparison between the observed CMD and four stellar isochrones at a richer metal content of [Fe/H]=$-0.4$.
The current comparison does not allow us to discriminate 
between the two adopted metal contents, due to the degeneracy along the RSGs sequence between age and metallicity.
Indeed, we note that considering a [Fe/H]=$-0.7$ the young population features are traced by isochrones with ages up to 100 Myrs, while at higher iron abundances ([Fe/H]=$-0.4$) they are justified by younger isochrones.
We also note that more quantitative constraints are partially hampered by the occurrence of differential 
reddening (\citetalias{Tantalo22}). The reddening vector plotted on 
the top right corner shows that the comparison might be affected by 
possible uncertainties in the mean reddening and/or in the distance 
modulus and/or in the mean adopted chemical composition.
More quantitative analysis would require accurate and deep 
optical--NIR--MIR photometry to improve the temperature/metallicity sensitivity and to mitigate 
the reddening uncertainty.

The comparison of the CMD of candidate young stars of NGC~6822 with models shows that the galaxy was still forming stars in a very recent epoch. This evidence is supported by the computed SFHs available in literature for NGC~6822 \citep[i.e.][]{Gallart96c, Cannon12} suggesting that the star formation activity increased over the whole main body of the galaxy in the last 100 Myr. Conversely, the SFH reconstructed by \citet{Fusco14} for six HST fields along the major axis of the galaxy revealed that the SFR of all the fields, but one, has slowly decreased in the last 1.5 Gyr.
Furthermore, \citet{Fusco14} derived the AMR for the last 5 Gyr, disclosing that 
the metallicity has grown with time from [Fe/H]=$-1.0$ to [Fe/H]=$-0.5$.
This range of values is compatible with the metallicities constrained from 
the comparison between the central regions of the galaxy and the isochrones.

\subsection{Intermediate stellar population} \label{sec:intermediate}

\begin{figure}[ht!]
  \centering
  \includegraphics[width=8.7cm]{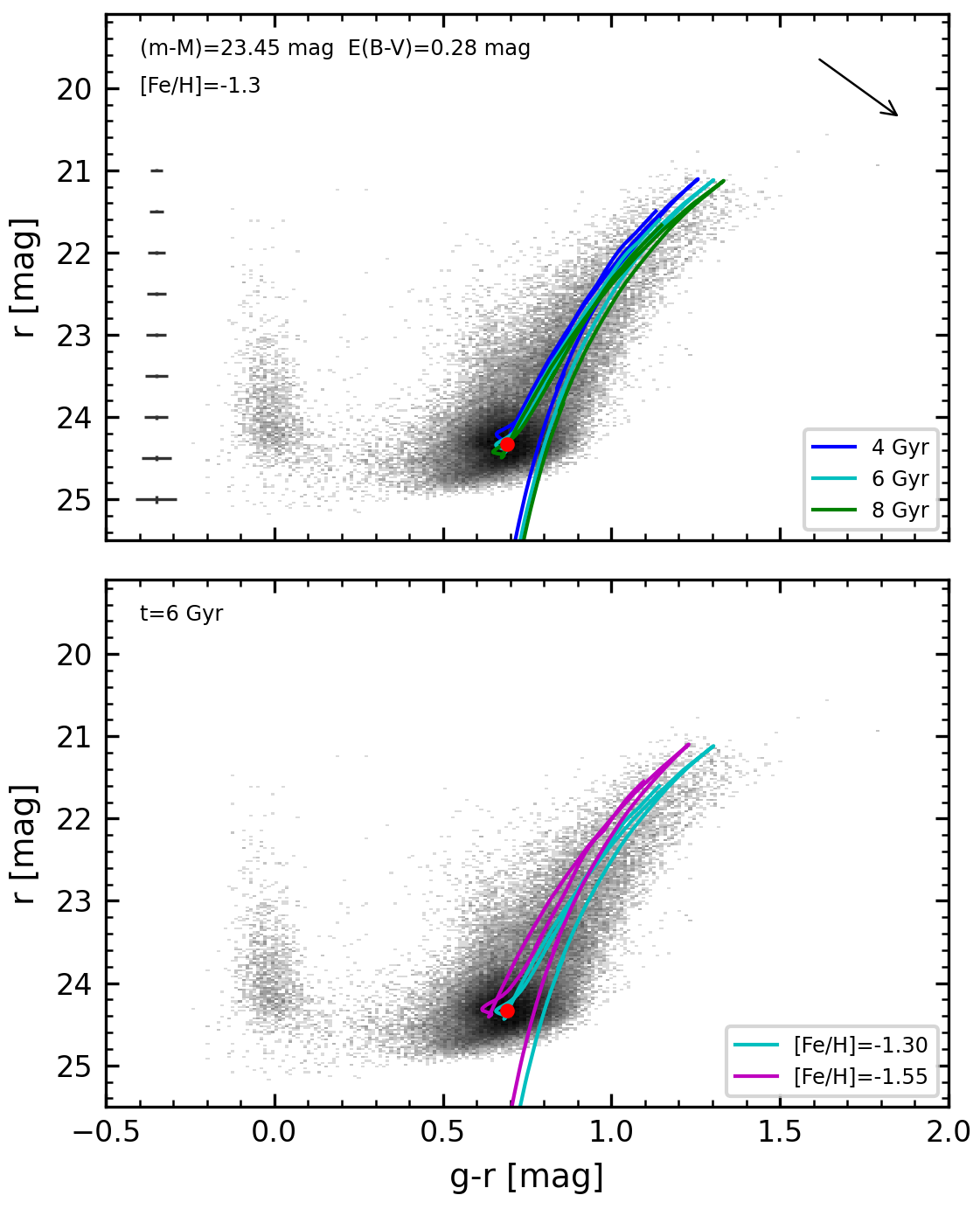}
  \caption{Top: same as the top panel of Figure~\ref{fig:CMDyoung}, but for the intermediate-age sample (see Figure~\ref{fig:map}). The three scaled-solar BaSTI-IAC  isochrones have the same metallicity ([Fe/H]=$-1.3$) and $4, 6$, and $8$ Gyr respectively (see labels).
           The red filled circle highlights the mean magnitude of the peak in the luminosity function of RC stars.
           The error bars plotted on the left side of the CMD display the intrinsic errors both in magnitude and in color (summed in quadrature).
           The black arrow plotted in the upper right corner shows the reddening vector for the extinction values we adopted.
           Bottom: same as the top panel, but the theoretical isochrones are for the same age (6 Gyr) 
           and different metallicities: [Fe/H]=$-1.3$ and $-1.55$. 
           \label{fig:interm}}
\end{figure}

The comparison between the scaled-solar isochrones and the observed $r$-($g-r$) CMD
for the intermediate-age population sample is shown in Figure~\ref{fig:interm}.
The catalog has been cleaned up using the same selection criteria on the 
$\chi$ and $Sharpness$ parameters as for the young sample, ending up with 75,831 stars.
The isochrones were plotted by using the same distance modulus and mean 
reddening adopted for the young population.

To compare theory and observations for 
the intermediate-age
stellar populations we employed isochrones with a metallicity chosen taking advantage of estimates based on  
C/M-metallicity relations for AGB stars, i.e.  
[Fe/H] = -1.3 $\pm$ 0.2 provided by \citet{Kacharov12}, 
[Fe/H] = -1.14 $\pm$ 0.07 by \citet{Sibbons12},
[Fe/H] = -1.38 $\pm$ 0.06 by \citet{Sibbons15},
[Fe/H] = -1.286 $\pm$ 0.095 by \citet{Hirsc20}  
and more recently [Fe/H] $\sim$ -1.25 ($\sigma$=0.04 dex) from \citetalias{Tantalo22}.

The top panel of the figure displays the observed CMD and three isochrones
for [Fe/H]=$-1.3$ and ages of $4, 6$ and $8$ Gyr. 
In the bottom panel we show isochrones
for an age of $6$ Gyr and different iron abundances ([Fe/H]=$-1.55$ and $-1.3$).
Note from the CMD that there are some YMS stars, because the intermediate-age sample was selected in a region of the galaxy where the disk of young stars is still present. 
To constrain the range of age and metallicity of the intermediate 
stellar population in NGC~6822, we considered as a reference the mean magnitude and color
of the peak of RC stars, at $r \sim$ 24.3 mag and $(g-r) \sim$ 0.7 mag and highlighted with a red filled circle. These have been derived by using a 3D histogram of the $i$-$(g-i)$ CMD. The CMD was split into a grid in magnitude and in color, while Z-axis shows the number of stars per grid point (see Figure~19 of \citetalias{Tantalo22}). The reader interested in a more detailed discussion concerning  the approach we adopted is referred to the Appendix~A of \citetalias{Tantalo22}.
The isochrone for t=6 Gyr and [Fe/H]=$-1.3$ properly fits the peak in magnitude of the observed RC. 
However, to account for the range in magnitude and 
color of the RC stars, isochrones with ages ranging from 4 to 8 Gyr and metal abundances ranging from [Fe/H]=$-1.3$ to $-1.55$  are required.
This shows that intermediate-age stellar populations experienced 
a modest chemical enrichment over an age range of at least 4-8 Gyrs. 
Determinations from SFHs suggested for this galaxy either a linear chemical enrichment law \citep{Gallart96b} 
or a broad variation according to the radial distance \citep[][see their Figure~9]{Fusco14}.

\subsubsection{HST CMDs for intermediate-age stars} \label{sec:HSTint}

\begin{figure*}[ht!]
  \centering
  \includegraphics[width=18cm]{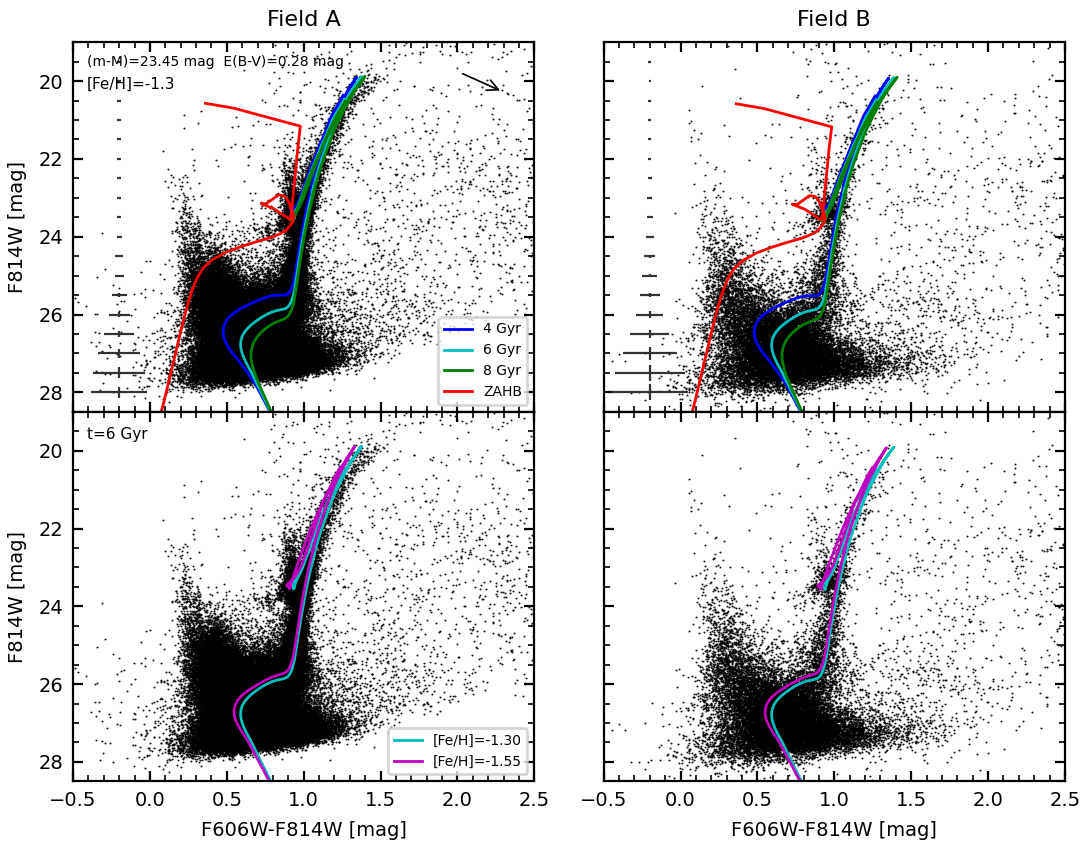}
  \caption{Top: comparison between the observed $F814W$-($F606W-F814W$) CMD of the HST field A (left) and field B (right) and the scaled-solar BaSTI-IAC isochrones of the top panel of Figure~\ref{fig:interm}, and the core He-burning sequence (ZAHB and RC) for the same metallicity (see labels).
  The error bars plotted on the left side of the CMD display the intrinsic errors both in magnitude and in color (summed in quadrature).
  The black arrow plotted in the upper right corner shows the reddening vector for the extinction values adopted in the current analysis.
  Bottom: same as the top panels, but the isochrones are for the same age (6 Gyr) and two different metallicities ([Fe/H]=$-1.3$ and $-1.55$).
  \label{fig:HSTint}}
\end{figure*}

Figure~\ref{fig:HSTint} shows the comparison between the observed $F814W$-($F606W-F814W$) 
CMD of the HST fields A (left panels) and B (right panels) 
and the same scaled-solar BaSTI-IAC isochrones superimposed on the 
$r$, $g-r$ CMD for the intermediate-age population adopted in Figure~\ref{fig:interm}.
The top panels also show, for a metallicity of [Fe/H]=$-1.3$, the 
core He-burning sequence (zero-age horizontal branch and RC).
The models were plotted by using the same distance modulus and mean reddening adopted in all these comparisons.
The extinction in the individual photometric bands was calculated by using the \citet{Cardelli89}'s reddening law
($A_{F606W}$=$0.96 A_V$, $A_{F814W}$=$0.61 A_V$).
The CMDs were cleaned up according to both their $Sharpness$ 
values ($-0.5 < Sharpness < 0.5$) and the photometric errors of the stellar objects 
($\sigma (F606W-F814W) < 0.2$ mag).

The RC stars can be easily identified in the CMD at 
$F814W$ between $23-24$ mag and $F606W-F814W$ colors between $0.9-1$ mag.
Moreover,  RGB stars display, as expected for high spatial resolution space-based 
observations, a narrower spread in color than the optical ground-based data.
The isochrones for ages in the range $4-8$ Gyr and 
with a metallicity of [Fe/H]=$-1.3$ (see the top panels of Figure~\ref{fig:HSTint}) are in fair 
agreement with the location of both the RC and the spread in color of RGB stars. 
Note, also, that the predicted core He-burning sequence properly fits both the 
magnitudes and the colors typical of RC stars.
In the bottom panels we overplotted isochrones with a fixed age of 
6~Gyr and metallicities of [Fe/H]=$-1.3$ and $-1.55$. The latter one 
fits quite well both the RC stars and the blue edge of the RGB. 
This means that the intermediate-age stellar populations likely have 
a more metal-poor tail. However, this age-metallicity degeneracy 
prevented us from constraining quantitatively their ranges.  
This evidence further strengthens the conclusions of our analysis on this population from the HSC@Subaru data.

\subsection{Old stellar population} \label{sec:old}

\begin{figure}[ht!]
  \centering
  \includegraphics[width=8.7cm]{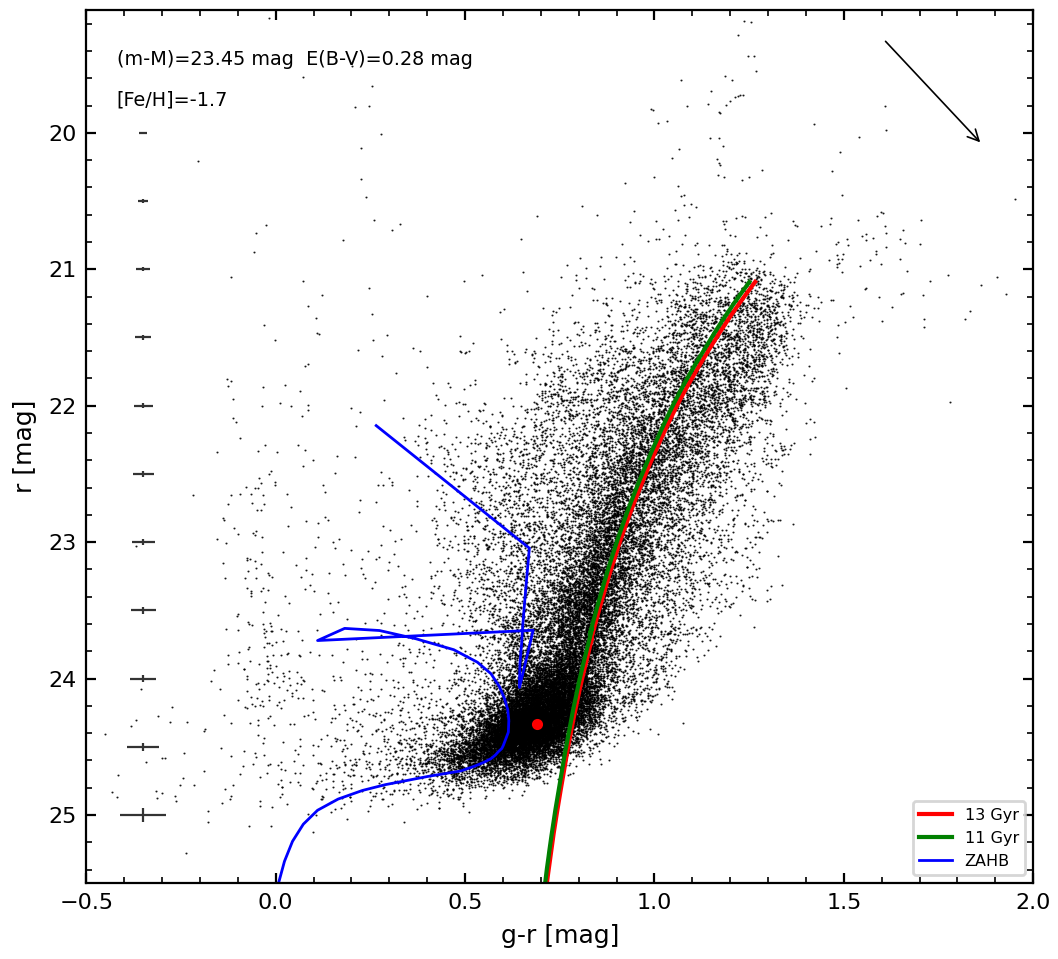}
  \caption{Comparison between the observed $r$-($g-r$) CMD of the old stellar population sample (see 
  Figure~\ref{fig:map}) and two scaled-solar BaSTI-IAC isochrones for ages equal to  $11$ and $13$ Gyr, and a core He-burning sequence. The metallicity has been fixed at [Fe/H]=$-1.7$. 
  The red filled circle highlights the mean magnitude of the peak in the luminosity function of RC stars, the same as in Figure~\ref{fig:interm}.
  The error bars plotted on the left side of the CMD display the intrinsic errors both in magnitude and in color (summed in quadrature).
  The black arrow plotted in the upper right corner shows the reddening vector for the extinction values adopted in our analysis.
           \label{fig:CMDold}}
\end{figure}

Figure~\ref{fig:CMDold} displays the comparison of the observed $r$-($g-r$) CMD
for the old population with two scaled-solar BaSTI-IAC isochrones and a core He-burning sequence.
Stellar objects were selected following criteria similar to the selection 
of young- and intermediate-age samples. However, we decided to add a further 
cut by using the photometric errors. 
We only retained sources with errors in color, calculated by summing in 
quadrature the errors on the individual magnitudes, smaller than 0.1 mag ($\sigma (g-r) < 0.1$ mag). 
The list has 34,928 stars with respect to the primary one.
This criterion and the large sample of galaxy stars allowed us to improve the cleaning 
of the CMD across the faint magnitude limit. 
The isochrones were plotted using the same distance modulus and mean reddening used for the young- and the intermediate-age sample. 
We superimposed onto the observed CMD isochrones with ages of $11$ (red line) and $13$ Gyr (green line)
and the core He-burning sequence (blue line), with a metallicity [Fe/H]=$-1.7$.
The adopted metallicity is based on spectroscopic measurements of individual RGB stars in NGC~6822 provided by 
\citet{Tolstoy01} ([Fe/H] = -1 $\pm$ 0.5) and by \citet{Kirby13} ([Fe/H] = -1.05 $\pm$ 0.01).
The adopted isochrones agree quite well with the magnitude range and the 
slope of the RGB in the CMD, thus suggesting a spread both in age and metallicity 
also for the old stellar population of NGC~6822.
Furthermore, the CMD shows evidence of a well-populated red horizontal branch (HB),
but the comparison with the core He-burning sequence discloses that the
photometric limit of our catalog does not allow us to possibly reveal
the presence of the very blue (hotter) HB stars. 
The adopted theoretical sequence for [Fe/H]=$-1.7$ traces
very well the lower limit of the red HB. Note that it attains, as expected 
for the selected ages and chemical composition, bluer colors when compared 
with RC stars (see the location on the CMD of the peak in magnitude of RC stars).

These qualitative results about the old population are compatible with 
the SFHs available in the literature. Indeed, \citet{Gallart96b} suggested  
that NGC~6822 has most likely started to form stars $15 - 12$ Gyr ago from a gas at 
low metallicity, around $Z_{i}=0.0001-0.0004$ which correspond to [Fe/H] = -2.2 $-$ -1.6. 
The metal abundances inferred from the isochrones are near the upper limit of this range. 
Moreover, \citet{Cannon12} found that the derived SFHs of six HST fields located 
along the major axis are qualitatively consistent with each other, suggesting that a 
high fraction of stars was formed between 14 and 6 Gyr ago.

The anonymous referee noted that we are suggesting an increase in iron abundance 
ranging from [Fe/H]=-0.7 to -0.4 for the young stellar population to 
[Fe/H]=-1.3/-1.5 for the intermediate-age stellar population and to [Fe/H]$\sim$-1.7 
for the old stellar populations. This evidence can be barely compared with current 
spectroscopic measurements, since we still lack detailed investigations concerning 
the iron distribution function among the different stellar components. 
Indeed, the few spectroscopic measurements available in the literature are from \citet{Venn01} using two RSG stars ([Fe/H]=-0.49$\pm$ 0.22), and from \citet{Kirby13} using RGB stars ([Fe/H]=-1.05$\pm$ 0.01).
However, the detailed SFH and age-metallicity relation (AMR) provided by \citet{Fusco14}
indicates that the iron abundance in the six HST pointings they analyzed is steadily increasing from [Fe/H]=-1.0/-1.3 ($\sim$5 Gyr ago) up to  [Fe/H]=-0.3/-0.6 in recent times (see their Figure~9).

When compared with similar investigations in the 
literature, our results show that the core He-burning sequence displays a well-defined 
continuum when moving from the horizontal portion of the ZAHB, where the He ignition  
is affected by electron degeneracy, to RC stars, whose He ignition is only partially affected 
by electron degeneracy, and to intermediate-mass stars 
who experience He ignition in non-degenerate conditions \citep[see, e.g.,][]{Salaris05}. 
The identification of these evolutionary properties further supports the evidence that NGC~6822 hosts a broad variety of core He-burning variable stars \citep{Pietr04,Baldacci05,Menni06,Feast12,Rich14}.
This means that NGC~6822 is a perfect 
laboratory to investigate advanced evolutionary phases, and in particular, to constrain the 
accuracy of old (RRLs), intermediate-age (ACs) and young (CCs) distance indicators.

\subsubsection{HST CMDs for old stars} \label{sec:HSTold}

\begin{figure*}[ht!]
  \centering
  \includegraphics[width=18cm]{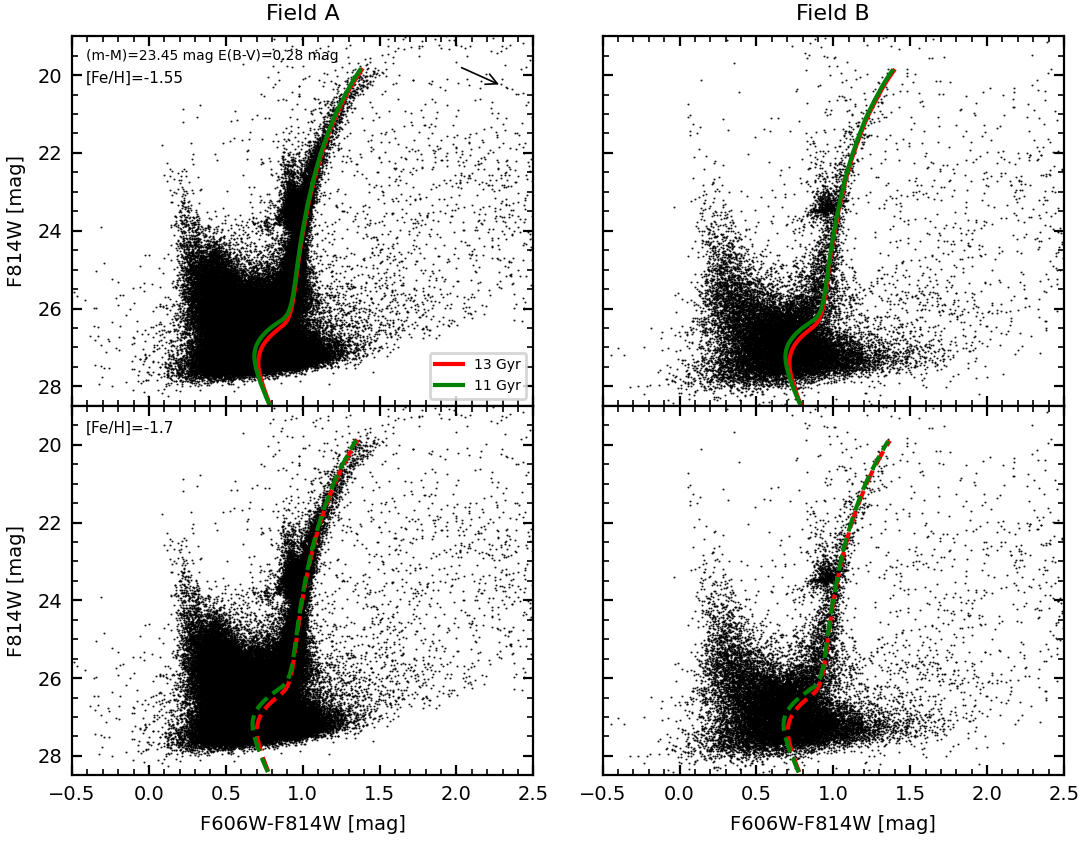}
  \caption{Top: comparison between the observed $F814W$-($F606W-F814W$) CMD of the HST fields A (left) and B (right) and the scaled-solar BaSTI-IAC isochrones with ages equal to 11 and 13 Gyr and [Fe/H]=$-1.55$ (see labels). The black arrow plotted in the upper right corner shows the reddening vector for the extinction values adopted in the current analysis.
  Bottom: same as the top panels, but the isochrones are for [Fe/H]=$-1.7$. 
           \label{fig:HSTold}}
\end{figure*}

\begin{figure}[ht!]
  \centering
  \includegraphics[width=8.8cm]{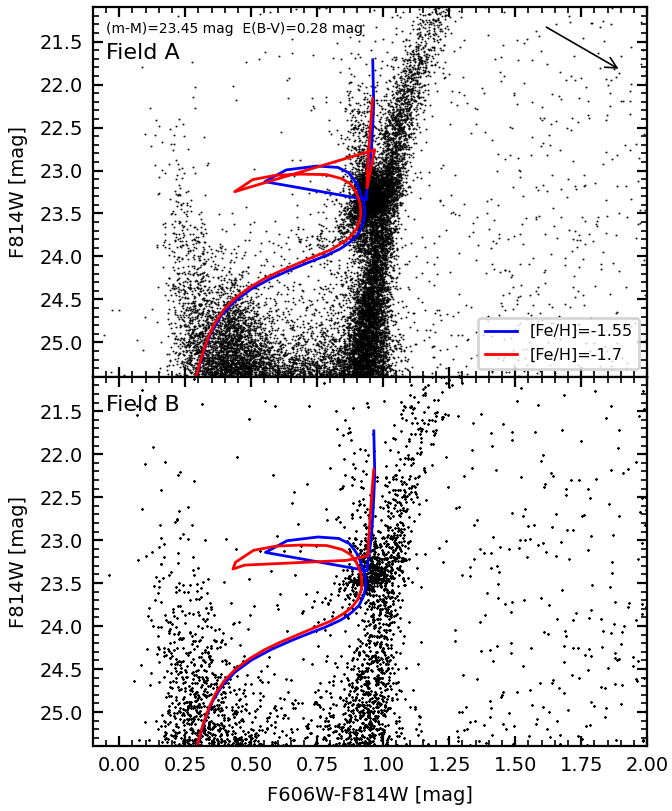}
  \caption{Top: comparison between the observed $F814W$-($F606W-F814W$) CMD of the HST field A and the core He-burning sequence for iron abundances [Fe/H]=$-1.55$ and [Fe/H]=$-1.7$ (see labels). The black arrow plotted in the upper right corner shows the reddening vector for the adopted extinction value.
  Bottom: same as the top panel, but for the HST field B.
           \label{fig:HSTzahb}}
\end{figure}

Figure~\ref{fig:HSTold}  shows the comparison between the observed 
$F814W$-($F606W-F814W$) CMD of the HST fields A (left panels) and B (right panels) 
and old scaled-solar BaSTI-IAC isochrones. 
The models were plotted by applying the same distance modulus and mean reddening previously adopted.
We already mentioned that the spread in color along the RGB is caused 
by variations either in age, or metallicity, or both. 
The adopted isochrones (11, 13 Gyr) with a fixed metal content ([Fe/H]=$-1.55$) 
takes account for magnitudes and colors of the stars distributed along the RGB 
(see top panels Figure~\ref{fig:HSTold}). However, we cannot determine accurately 
the iron abundance(s) of the old stellar population, since the two isochrones 
plotted in the bottom panels of Figure~\ref{fig:HSTold} with the same ages and 
a slightly more metal-poor iron abundance ([Fe/H]=$-1.7$), still provide a good 
fit of the stars distributed along the RGB. 

Figure~\ref{fig:HSTzahb} shows 
the comparison between the observed HB stars in fields A (top) 
and B (bottom) and the predicted He-burning sequence 
for the same iron abundances adopted in Figure~\ref{fig:HSTold}. 
The CMDs were zoomed onto the region between the old HB and RC stars. 
The photometric accuracy of the HST photometry allows us to easily identify 
a group of old red HB stars located at $F814W \sim 24$ mag and 
$F606W$-$F814W\sim0.80$ mag, together with a number of candidate RRL 
stars with bluer colors and a larger spread in magnitude 
located at $F814W \sim 24$ mag and $F606W$-$F814W\sim0.50$ - $0.90$ mag.  
Time series multi-band photometric data and He-burning variable stars 
will be investigated in a forthcoming paper. 
A firm identification of blue HB stars is hampered by the presence of a 
significant number of YMS stars that overlap in the same CMD region.
The HST photometry further supports the continuum along the core  
He-burning sequence brought forward by the ground-based wide field 
$gri$ photometry.

\subsection{Identification of the AGB clump} \label{sec:delta}

\begin{figure}[ht!]
  \centering
  \includegraphics[width=9cm]{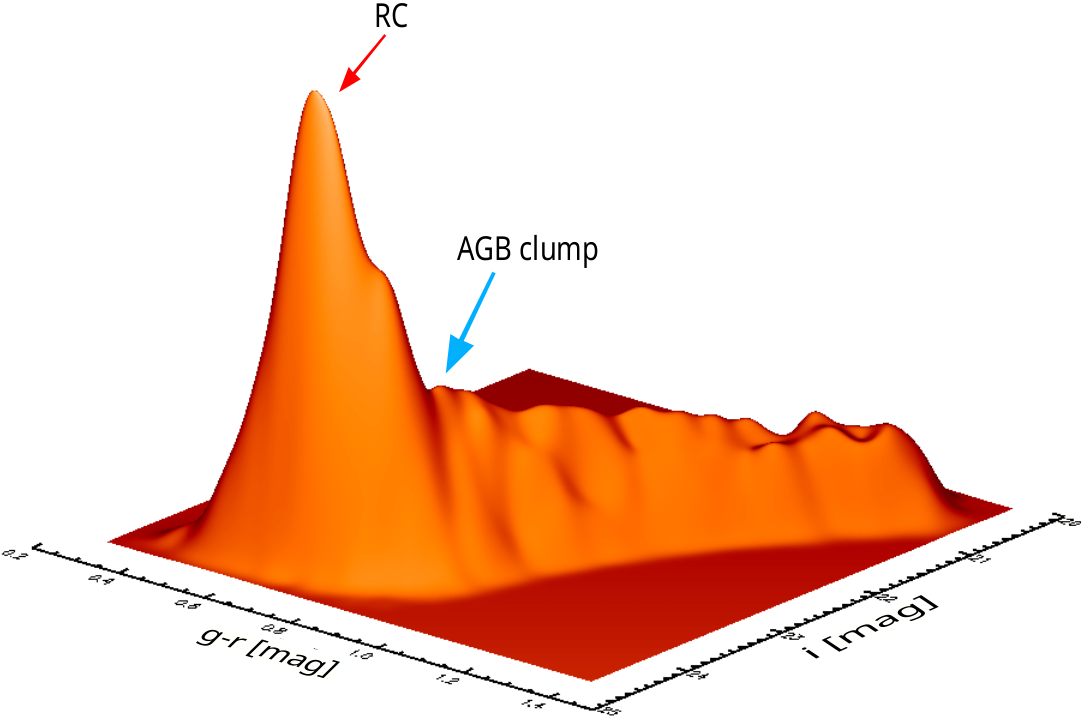} 
  \includegraphics[width=8.7cm]{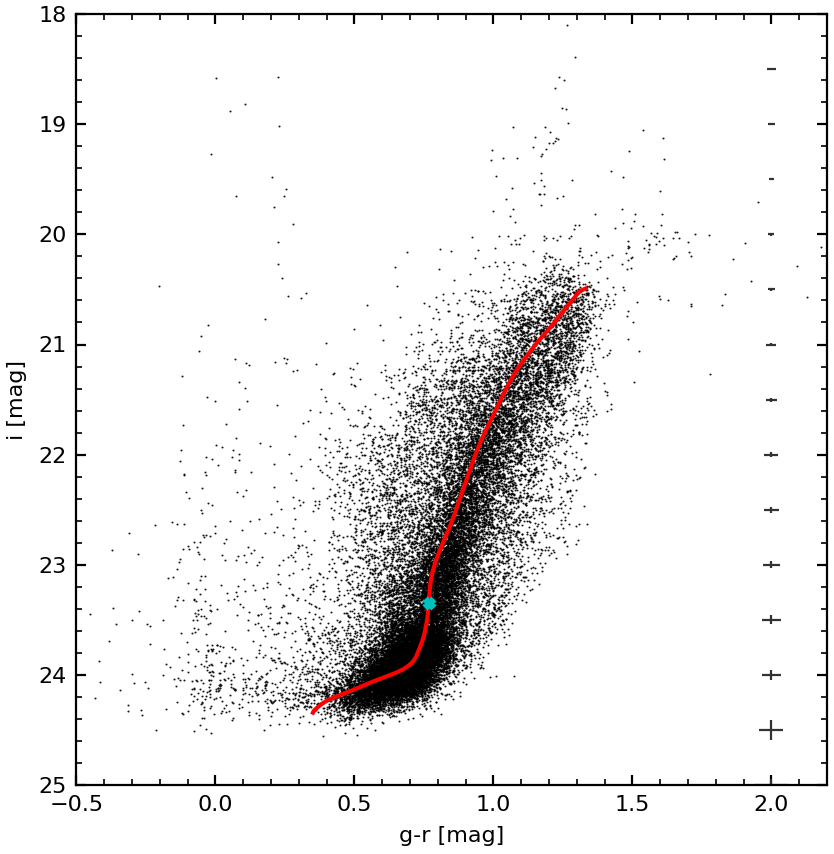}
  \caption{
  Top: three-dimensional CMD of RG stars. X-axis, color $g-r$; y-axis, magnitude $i$; 
  z-axis, luminosity function (arbitrary units).
  The red and cyan arrows mark the main peak associated to the RC stars and to 
the AGB clump.
  Bottom: Ridge line in the observed $i$-($g-r$) CMD (red solid line) used to derive the luminosity function of the RG stars. Our identification of the AGB clump is highlighted as a cyan filled cross. 
  The error bars plotted on the right side of the CMD display the intrinsic errors both in magnitude and in color (summed in quadrature).\label{fig:ridge}}
\end{figure}

\begin{figure}[ht!]
  \centering
  \includegraphics[width=8.7cm]{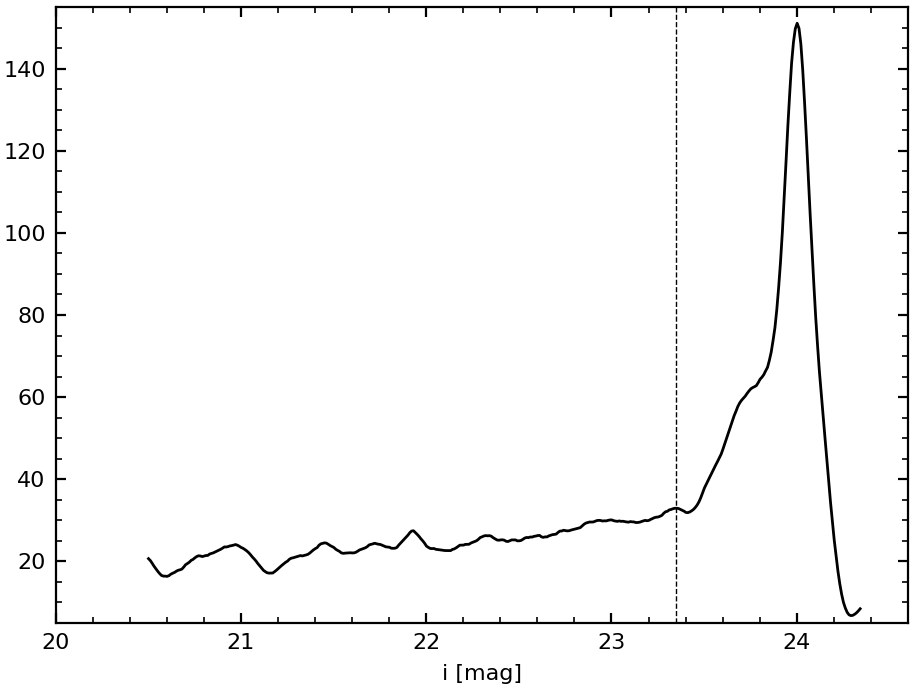}
  \caption{Marginal (arbitrary units) based on the ridge line of the RG stars versus $i$-band magnitude. The vertical dashed line marks the peak of the AGB clump. 
           \label{fig:profile}}
\end{figure}

\begin{figure}[ht!]
  \centering
  \includegraphics[width=8.7cm]{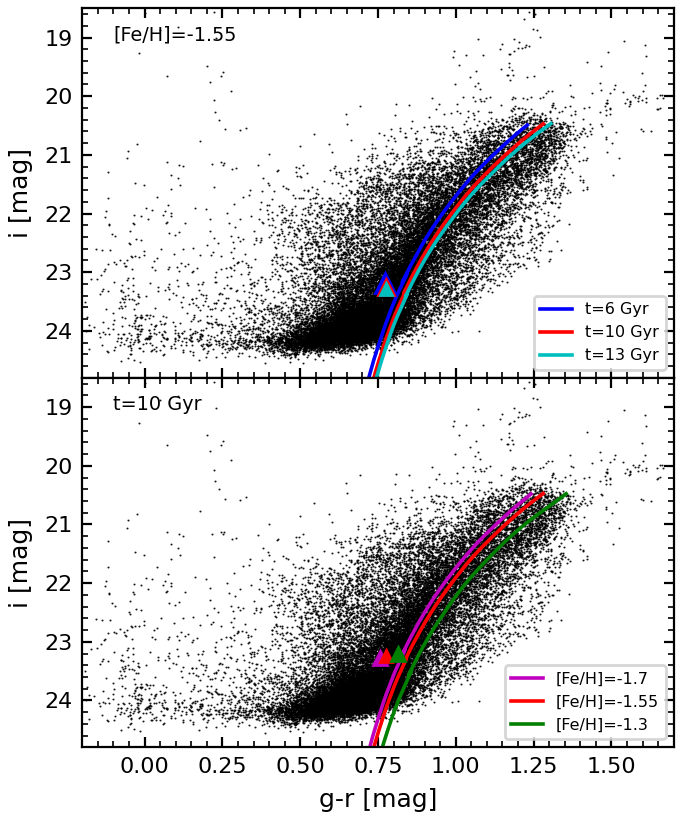}
  \caption{Top: $i$-($g-r$) CMD of the old stellar population sample together 
  with three scaled-solar BaSTI-IAC isochrones for a fixed metallicity ([Fe/H]=$-1.55$) 
  and stellar ages of 6, 10 and 13 Gyr (see labels). Only the evolutionary phases until  
  the TRGB are shown here. For each isochrone the filled triangles denote the position of the ignition of the double shell burning, typically associated 
  to the AGB clump. Bottom: same as the top panel, but the isochrones were computed at fixed age 
  (t=10 Gyr) and different metal contents of $-1.7$, $-1.55$, and $-1.3$ (see labels).
           \label{fig:AGBclump}}
\end{figure}

Based on the recent investigation of NGC~6822 stellar populations by \citet{Nally24} 
and their identification of the AGB clump in the MIR ($F200W\sim 21$ mag, 
$F115W-F200W\sim 0.8$ mag) CMD, we decided to take 
advantage of the current accurate and deep optical catalog to identify the same 
evolutionary phase, and to perform a detailed comparison with theoretical models. 
Dating back to \citet{Pulone92a, Pulone92b} it was recognized that the AGB clump 
(the ignition of double shell burning) is 
indeed a good standard candle, minimally affected by changes in iron 
abundance and stellar age. The same evidence was also supported by 
observations of Galactic GCs \citep{Ferraro92}.  

To properly identify the AGB clump we derived a three-dimensional (3D) CMD that includes the 
color $g-r$, the magnitude $i$, and the corresponding luminosity function of RG stars.
To overcome subtle problems caused by the binning of the data, 
we smoothed the 3D distribution associating to the position of each star 
in the CMD 
a Gaussian with unitary weight and sigma equal to the error in the $g-r$ color 
measurement. The global 3D CMD  (see the top panel of  Figure~\ref{fig:ridge}) 
was computed by integrating all the Gaussians associated to  individual stars 
over the entire color and magnitude range of RG stars. 
This 3D CMD allowed us to measure the ridge line (variation of the main 
peaks) of RG stars in NGC~6822 as a function of magnitude and color. 
The bottom panel of Figure~\ref{fig:ridge} shows the observed $i$-($g-r$) 
CMD of the old stellar population with the ridge line derived 
on the basis of 3D CMD.

Furthermore, to trace the variation of the luminosity function of red giant 
stars we computed the marginal of their distribution. This means that individual 
RGs were shifted in magnitude and color to their closest point 
along the ridge line. Subsequently, we estimated their global luminosity function 
by smoothing the distribution with a Gaussian kernel. Figure~\ref{fig:profile} 
displays the marginal (arbitrary units) in the $i$-band magnitude ranging from 
the tip of the RGB down to the RC. The marginal is characterised by a main peak at 
$i\sim 24$ mag associated to RC stars (see Appendix~A of \citetalias{Tantalo22}), 
followed by a few secondary peaks at fainter magnitudes. Based on the results 
provided by \citet{Nally24}, according to which the AGB clump is roughly located 
1 mag brighter than the RC, we identified the peak at $i\sim 23.35$ mag with 
the AGB clump and the vertical dashed line plotted in this figure marks its location. 
The current identification is also highlighted with a cyan arrow on the 3D CMD in the top panel of Figure~\ref{fig:ridge},
on which we have also indicated the main peak of RC stars with a red arrow,
and as a cyan filled cross on the $i$-($g-r$) CMD in the bottom panel of Figure~\ref{fig:ridge}.

To validate this identification, we compared the observations with several stellar 
isochrones from the BaSTI database, applying the same distance modulus and reddening 
adopted in the previous sections.
In particular, we checked the location in the $i$-($g-r$) CMD of the point marking 
the exhaustion of core He-burning and the ensuing double shell (hydrogen, helium) 
burning. This evolutionary phase is typically associated with  the AGB clump 
\citep{Salaris05}.

The two panels in Figure~\ref{fig:AGBclump} show the comparison between 
the $i$-($g-r$) CMD of the old stellar population and isochrones at fixed metallicity 
([Fe/H]=$-1.55$) and stellar ages of 6, 10 and 13 Gyr (top panel), 
and isochrones at fixed stellar age (t=10 Gyr) and metallicities equal to $-1.7$, $-1.55$, and $-1.3$ (bottom panel). 
In this case we only selected the portion of the isochrones till the tip of the RGB. 
For each isochrone a filled triangle denotes the position of the corresponding 
AGB clump. Their location on the CMD agrees very well with the magnitude and color 
of the observed AGB clump ($i\sim 23.35$ mag and $g-r\sim 0.77$ mag; see Figure~\ref{fig:ridge}). 
The models also show just marginal variations in magnitude 
and color for large variations of stellar age ($6-13$ Gyr) and metal content 
([Fe/H]=$-1.7/-1.3$).

\section{AGB stellar population} \label{sec:agb}

Several authors investigated on the identification of O- and C-rich AGB stars 
in complex stellar systems, leading to the development of several different   
photometric and spectroscopic diagnostics. Some studies have used a 
CC-D based on a mix of broad- (R,I) and narrow-band (CN, TiO) photometry 
\citep[hereinafter~\citetalias{Letarte02}]{Albert00, Battinelli00, Letarte02}, 
and this approach has been established to be reliable to select C-type stars.
Others took advantage of NIR CMDs and CC-Ds 
\citep[][]{Kang06,Sibbons12,Ren21,Li25}, and a combination of NIR and MIR 
CMDs \citep[see e.g.][]{Boyer11,Hirsc20}. In a few cases, low-resolution spectroscopy 
helped to improve the NIR photometric selection criteria \citep{Kacharov12,Sibbons15}.
More recently, the use of NIR and MIR bands has been tested with the filters mounted 
on the NIRCam and MIRI detectors on board the JWST \citep{Nally24,Boyer24}, 
showing very interesting results. 

In our previous work, we have investigated in great detail the AGB stellar population in NGC~6822. We decided to mainly focus on the use of the CC-Ds because the CMDs in the optical, NIR and MIR bands don't allow us to remove the degeneracy in color between the two AGB subgroups, as the C-type stars at fainter magnitudes tend to have similar colors to the O-type.
We first adopted approaches suggested in the literature, such as NIR and $(Cn-TiO)$-$(R-I)$ CC-Ds. The former method, as described by \citet{Kang06}, is based on the use of the histogram in NIR colors ($J-K$, $H-K$) to distinguish the two star samples, but the CC-Ds display a well-defined color sequence and don't solve the degeneracy. The latter method, used by \citetalias{Letarte02}, has been proved to be very solid in the identification of the C-rich stars, whereas the O-rich stars are hampered by the contamination of field red giants.
Therefore, we defined new solid diagnostics using several new optical-NIR-MIR CC-Ds, on which the O- and C-rich stars are distributed along either two almost parallel sequences (i.e. on $i-[3.6]$-$g-i$, $i-[4.5]$-$g-i$, $i-[3.6]$-$r-i$, $i-[4.5]$-$r-i$, $r-K$-$r-i$ and $i-K$-$r-i$ CC-Ds) or sequences with different slopes (i.e. on $i-[3.6]$-$r-J$ and $i-[4.5]$-$r-J$ CC-Ds).
In this paper we focus our attention on the comparison between predictions based on 
stellar evolution models and observed CMDs and CC-Ds of the galaxy AGB population.

\begin{figure}[ht!]
  \centering
  \includegraphics[width=8.7cm]{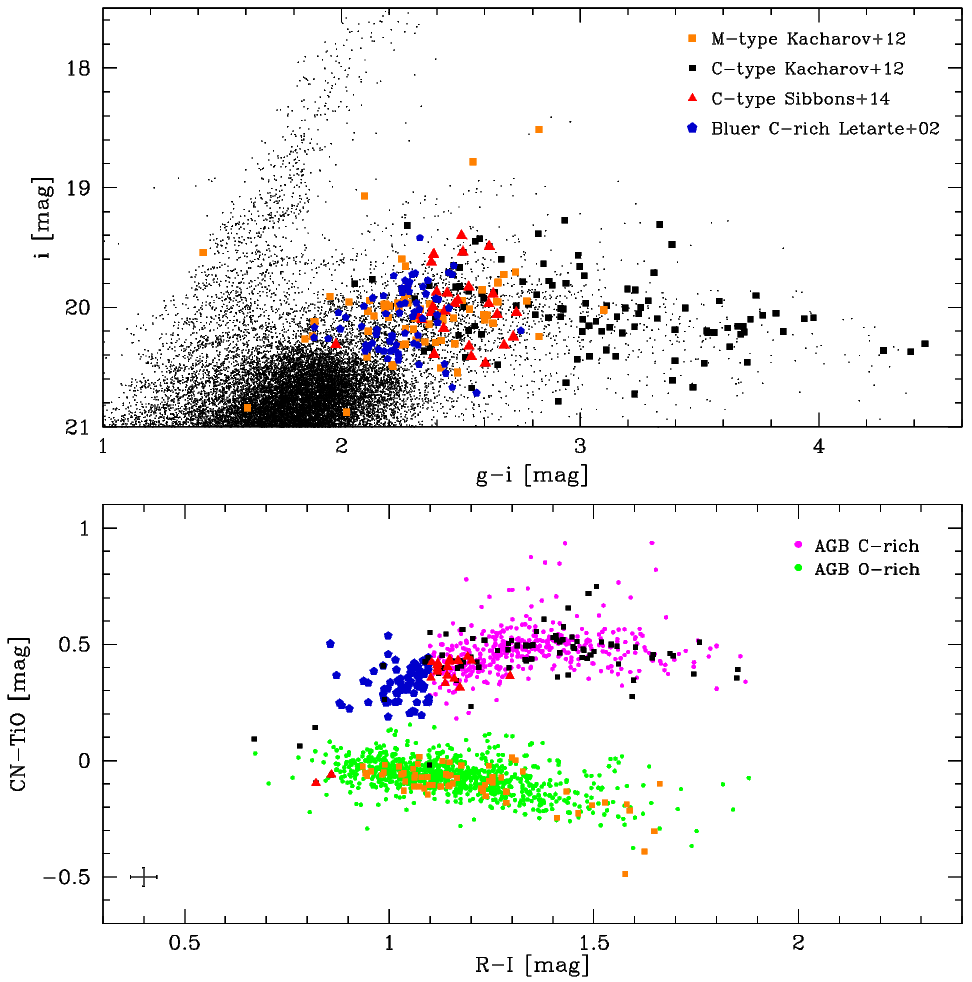}
  \caption{Top: $i$-($g-i$) CMD of the candidate galaxy stars in NGC~6822.
           The CMD is zoomed on the bright portion to show the region where AGB stars are located. 
           Black squares and red triangles display spectroscopically confirmed C-type stars
           \citep{Kacharov12, Sibbons15}, orange squares show confirmed M-type stars \citep{Kacharov12} 
           and blue pentagons mark the `bluer' C-rich stars, as defined by \citetalias{Letarte02} 
           (see bottom panel).
           Bottom: $(Cn-TiO)$-$(R-I)$ CC-D for candidate AGB stars 
           in NGC~6822. The diagram is based on multi-band photometry collected by 
           \citetalias{Letarte02} and shows a clear separation between candidate C-rich (magenta dots) and O-rich (green dots) stars. Symbols and colors of the spectroscopic sample (C-, M-type) and the `bluer' C-rich stars are the same as in the top panel.
           The error bars plotted on the left-bottom side of the CC-D display the intrinsic errors in colors (summed in quadrature).
           \label{fig:CMDagb}}
\end{figure}

The diagrams in Figure~\ref{fig:CMDagb} are the same as Figure~12 of \citetalias{Tantalo22}. 
The top panel is a zoom-in on the bright 
portion of the $i$-($g-i$) CMD to highlight our selection of the AGB stars,
on which we over-plotted the spectroscopic samples of C- and M-type stars 
available in the literature\footnote{The reader interested in more details 
concerning the spectroscopy of AGB stars is referred to Section~5.2 of 
\citetalias{Tantalo22}} for NGC~6822.  
The spectroscopic identification of C-type stars \citep{Kacharov12,Sibbons15} 
and M-type stars \citep{Kacharov12} are marked with different symbols. 
The blue pentagons display the so-called `bluer' C-rich stars as defined 
by \citetalias{Letarte02} and discussed in the next paragraph.
The bottom panel of the figure is the $(Cn-TiO)$-$(R-I)$ CC-D used by \citetalias{Letarte02} to select their sample of C- and O-rich stars, exploiting multiband photometry (CN, TiO, R and I) collected with both the Swope telescope and the wide-ﬁeld imager CFH12K at CFHT.
The plot shows the cross-match between our AGB sample and the catalog by \citetalias{Letarte02}.
The \lq{bluer\rq} C-rich stars are the same as the top panel, and 
have been defined by \citetalias{Letarte02} on this CC-D as
the stars that are systematically bluer than the $(R-I)$ color cut 
typically used to define the \lq{canonical\rq} C-rich stars ($(R-I)=$ 1.1 mag).
To perform a homogeneous comparison with similar results available in the 
literature, \citetalias{Letarte02} decided to exclude them from their sample 
of C stars.

\begin{figure*}[ht!]
  \centering
  \includegraphics[width=18cm]{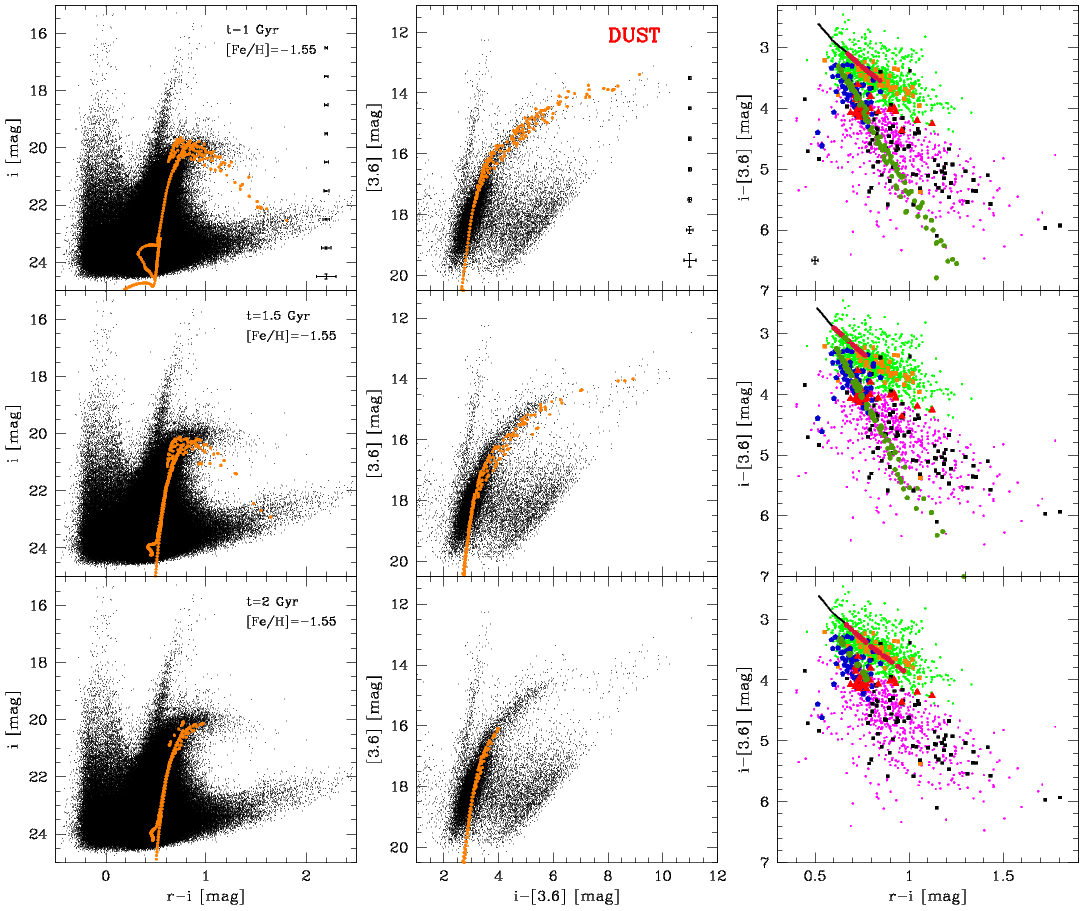}
  \caption{Left: comparison between the observed $i$-($r-i$) CMD of the candidate galaxy stars 
           and three PARSEC+COLIBRI stellar isochrones for a metallicity of [Fe/H]=$-1.55$ 
           and different assumptions about age: 1 Gyr (top), 1.5 Gyr (middle) and 2 Gyr (bottom). 
           They have been corrected for a true distance modulus of $23.45$~mag and a mean reddening $E(B-V)=0.28$~mag.
           The isochrones have been computed considering a dust composition of 
           60\% Silicate and 40\% AlOx for M stars and of 85\% AMC and 15\% SiC for C stars.
           The error bars plotted on the right side of the CMD display the intrinsic errors both in magnitude and in color (summed in quadrature).
           Middle: same as the left panels, but for the optical-MIR $[3.6]$-($i-[3.6]$) CMD.
           The MIR photometry ($[3.6]$ magnitudes) is from \citet{Khan15}.
           Right: same as the left panels, but for the optical-MIR ($i-[3.6]$)-($r-i$) CC-D. 
           The magenta and green dots display, respectively, the photometric samples of the AGB C- and O-rich stars. Symbols and colors of the spectroscopic samples (C- and M-type) and 
           of the `bluer' C-rich stars are the same as in Figure~\ref{fig:CMDagb}.
           For the sake of clarity, only the AGB evolutionary stage predicted by the models is highlighted here. 
           The isochrone section plotted as a black solid line show the E-AGB phase, 
           while that plotted as dots display the TP-AGB phase. The crimson dots are the isochrone portion populated by objects with an abundance ratio C/O$<$1, while the dark green dots correspond to C/O$>$1.
           The error bars plotted on the left-bottom side of the CC-D (top) display the intrinsic errors in colors (summed in quadrature).
           \label{fig:isoage}}
\end{figure*}

\begin{figure*}[ht!]
  \centering
  \includegraphics[width=18cm]{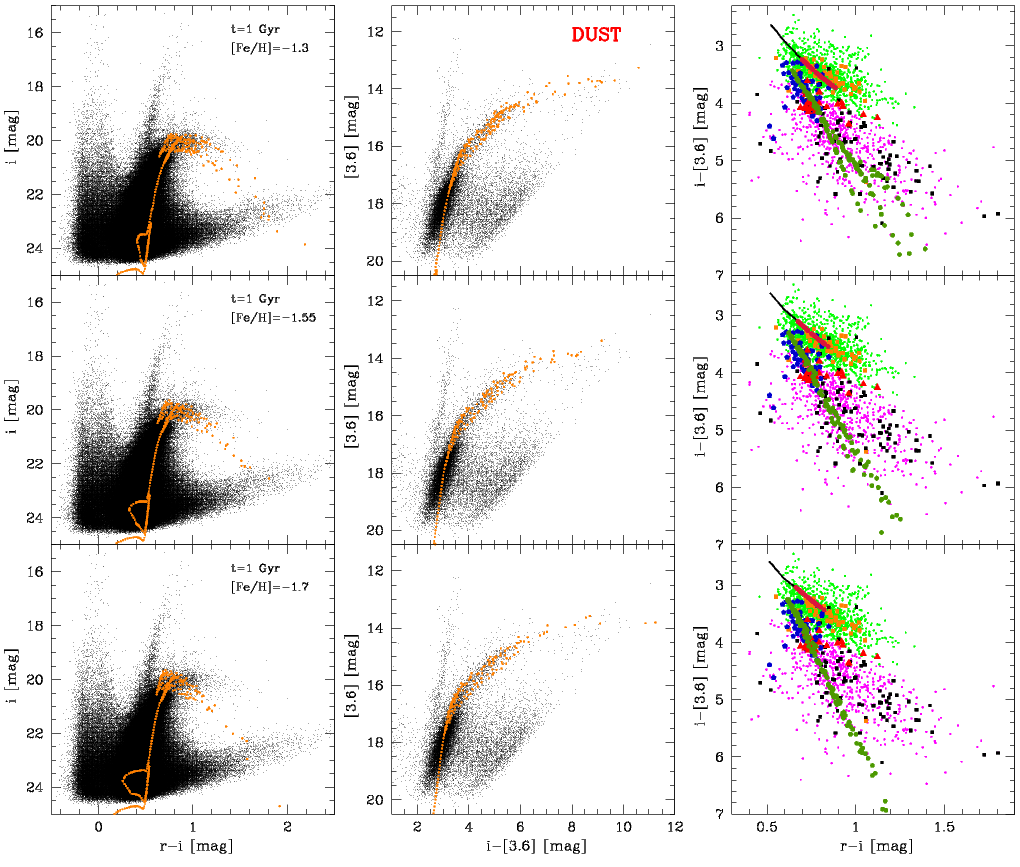}
  \caption{Left: same as the left panels of Figure~\ref{fig:isoage}, but the PARSEC+COLIBRI stellar 
           isochrones are for the same age of 1 Gyr and different assumptions about metallicity: 
           [Fe/H]=$-1.3$ (top), [Fe/H]=$-1.55$ (middle) and [Fe/H]=$-1.7$ (bottom). 
           Middle: same as the left panels, but for the optical-MIR $[3.6]$-($i-[3.6]$) CMD. 
           Right: same as the right panels of Figure~\ref{fig:isoage}, but for the same isochrones of the left panels. 
           \label{fig:isomet}}
\end{figure*}

\begin{figure*}[ht!]
  \centering
  \includegraphics[width=18cm]{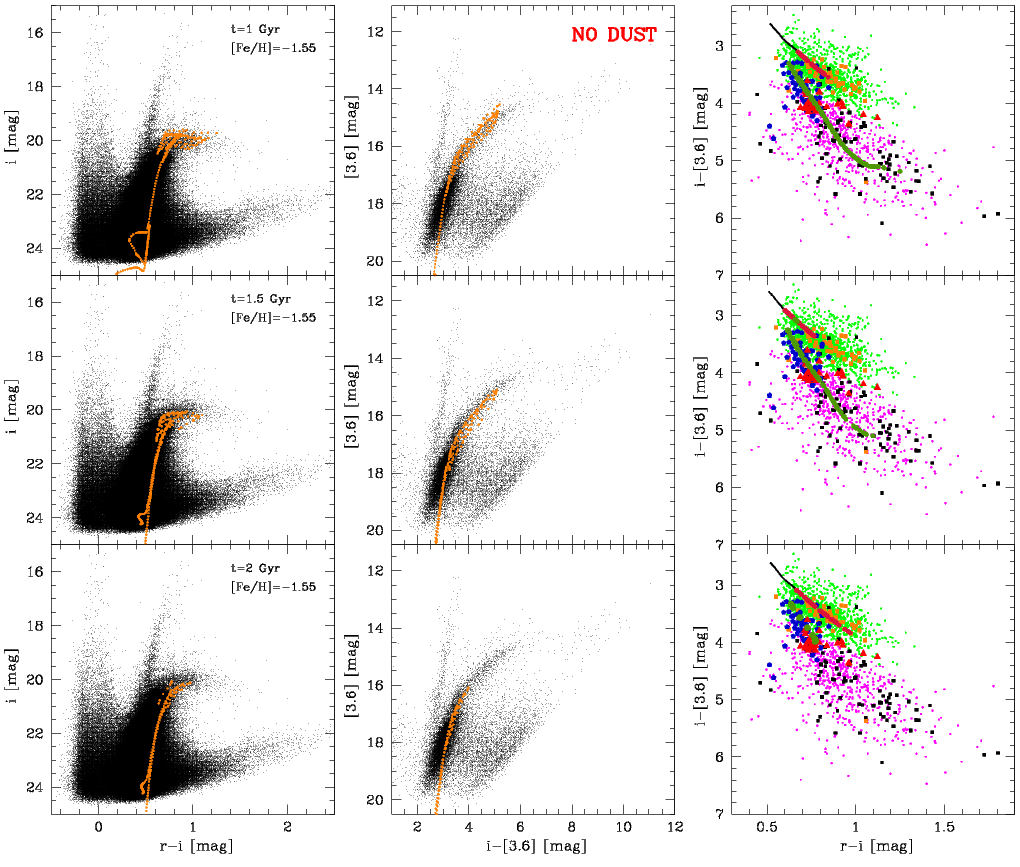}
  \caption{Same as Figure~\ref{fig:isoage}, but the isochrones have been computed without including the effect of circumstellar dust for M and C stars.
           \label{fig:ageNodust}}
\end{figure*}

\begin{figure*}[ht!]
  \centering
  \includegraphics[width=18cm]{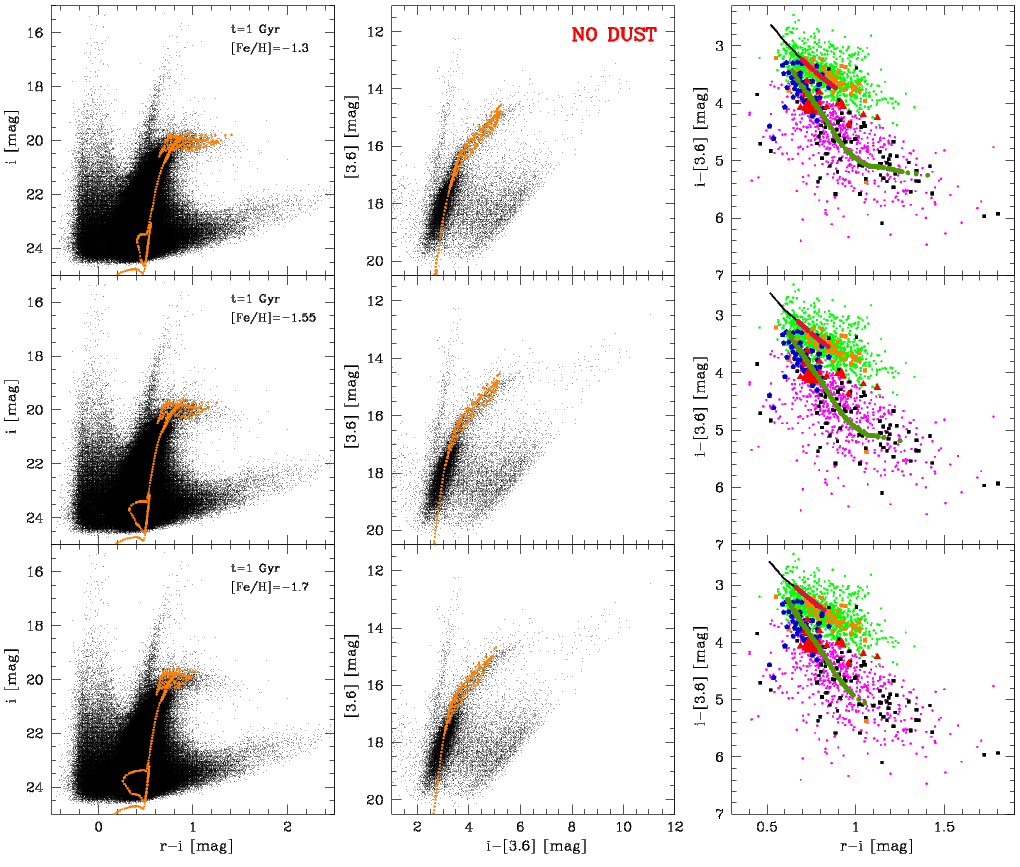}
  \caption{Same as Figure~\ref{fig:isomet}, but the isochrones have been computed without including the effect of circumstellar dust 
  for M and C stars. 
           \label{fig:metNodust}}
\end{figure*}

To constrain the age and metallicity ranges of the AGB stellar population,
we compared both the CMDs and the CC-Ds 
with the PARSEC+COLIBRI stellar isochrones \citep{Bressan12,Marigo13} retrieved from the CMD 3.7 web interface.
The results that we are going to discuss are shown in Figs.~\ref{fig:isoage},
\ref{fig:isomet}, \ref{fig:ageNodust} and \ref{fig:metNodust}.
The comparison with each isochrone was made by using both
the optical $i$-($r-i$) and the optical-MIR $[3.6]$-($i-[3.6]$) CMDs,
and the optical-MIR ($i-[3.6]$)-($r-i$) CC-D. Among all the CC-Ds we have defined
in \citetalias{Tantalo22}, we chose the latter one for reasons that will become
clearer in the following discussion.
Furthermore, we selected the isochrones taking into account both the presence
of circumstellar dust in stars during the TP-AGB phase
and its absence.

Figure~\ref{fig:isoage} shows the results of the comparison between the aforementioned
CMDs and CC-D and three PARSEC+COLIBRI stellar isochrones for a fixed metallicity of [Fe/H]=$-1.55$
and ages of 1 Gyr (top panels), 1.5 Gyr (middle panels) and 2 Gyr (bottom panels).
The circumstellar dust during the TP-AGB phase has been considered with
a dust composition of 60\% Silicate and 40\% Aluminium Oxide (AlOx) for M-type stars
and of 85\% Amorphous Carbon (AMC) and 15\% Silicon Carbide (SiC) for C-type stars, as suggested in the CMD 3.7 web interface.
The isochrones were adjusted to the same distance modulus ($23.45$ mag) and mean reddening ($E(B-V)=0.28$ mag) adopted in the previous analysis of the young, intermediate and old components.
A glance at the optical CMDs clearly shows that the total range in magnitude and color
covered by AGB stars is reproduced fairly well by the \lq{dusty\rq} isochrone at 1 Gyr.
However, the slope of the theoretical TP-AGB phase is steeper than the observed one. 
On the other hand, the same isochrone superimposed on the optical-MIR CMD is in very good agreement 
with both the slope and the whole extension of the AGB evolutionary sequence. 
This means that the discrepancy seen in the optical CMD at very red colors may 
be due to inaccurate bolometric corrections, causing a too steep increase in the 
magnitudes compared to the data. 
Indeed, the presence of the dust-enhanced AGB  stars require very large bolometric 
corrections for the the optical filters, compared to the IR. The same effect can 
also be seen in the isochrones with different ages.
The 1.5 Gyr isochrone in both CMDs covers the entire range in color of the AGB stars 
but at slightly larger magnitudes, while the isochrone at 2 Gyr covers a smaller 
interval of colors than observed, i.e. 0.2 mag over an observed range of 1 mag in 
$g-i$ and 1 mag over an observed range of 4 mag in $i-[3.6]$.
We point out here that during our analysis we made several tests by changing the dust composition for the M- and C-type stars among those provided in the CMD 3.7 web interface for the COLIBRI isochrones. We found that, independently of its composition, circumstellar dust always causes the same effects in the optical $gri$ filters. For these reasons, we have reported here only the results obtained by using the dust composition described before.

It is worth making a separate discussion on the comparison between the optical-MIR
CC-D and the three isochrones, shown in the right panels of Figure~\ref{fig:isoage}. 
For the sake of clarity, the isochrones only show the AGB evolutionary phases,  
with the early-AGB (E-AGB) and TP-AGB denoted by different symbols. 
These diagrams point out a very interesting result: for the first time it is clear 
that on the optical-MIR ($i-[3.6]$)-($r-i$) CC-D the theoretical M- and C-type tracks 
are on two distinct color sequences, fully consistent with those defined by the 
spectroscopic and photometric samples. This evidence also support 
the use of the new photometric diagnostics developed in \citetalias{Tantalo22} 
to identify and characterize O- and C-rich stars in stellar systems.
Further evidence is that the colors, hence the location, of the `bluer' C-rich
are consistent with the typical optical-MIR colors of the observed C-rich sample
and they are well matched by the C-type sequence of the isochrones. For this reason, 
in contrast with \citetalias{Letarte02}, we included them in the 
sample of \lq{canonical\rq} C-type AGB stars. 
The three selected isochrones properly match  the range in magnitude
and color of bright AGB stars, suggesting that they cover, at fixed metal content 
([Fe/H]=$-1.55$), ages ranging from 1 to 2 Gyr. Note that bright AGB stars allow us 
to investigate stellar populations systematically younger than RC stars 
(discussed in Section~\ref{sec:intermediate}). 
It is worth mentioning that the current sample also includes AGB stars from 
older stellar populations' progenitors, but we are not able to disentangle them
from younger AGB stars. Therefore the quoted range should be considered as a 
lower limit to the age range covered by bright AGB stars.

Figure~\ref{fig:isomet} displays the same plots of Figure~\ref{fig:isoage}, but the three
PARSEC+COLIBRI isochrones are for a fixed age of 1 Gyr and 
[Fe/H]=$-1.3$ (top panels), [Fe/H]=$-1.55$ (middle panels), and [Fe/H]=$-1.7$ (bottom panels).
The results of this comparison are similar to the previous one. 
In the optical-MIR CMD the three isochrones are a good match to the
slope and the location of the observed AGB, while in the optical CMD the slope 
disagrees with the observations. As for the optical-MIR CC-D, the evidence is the 
same as in Figure~\ref{fig:isoage} and the isochrones appear to reproduce a good 
fraction of the range in color covered by both O-rich and C-rich samples.
Our results suggest that intermediate-age 
stellar populations, including both RC (see Section~\ref{sec:intermediate}) and bright AGB stars, have ages ranging from 
$\sim$1 to $\sim$8 Gyr and metal abundances between [Fe/H]=$-1.3$ and [Fe/H]=$-1.7$.
This further supports the modest chemical enrichment experienced by this galaxy 
when moving from old to intermediate-age stellar populations.

Figures~\ref{fig:ageNodust} and \ref{fig:metNodust} are the same as  
Figures~\ref{fig:isoage} and \ref{fig:isomet}, but we used PARSEC+COLIBRI isochrones 
without including the effect of circumstellar dust for M- and C-type stars.
In this case   
the slope of TP-AGB phase models is in good agreement with the observed one 
in both the optical and optical-MIR CMDs. However, the models cover a smaller 
color range than observed. Indeed, due to the absence of dust, the bolometric corrections for the optical filters are greatly reduced.
As for the optical-MIR ($i-[3.6]$)-($r-i$) CC-D,
the absence of the dust does not affect the location or the slope of the theoretical M-type sequence, on the contrary the C-type sequence shows a change in the slope. 

\begin{figure*}[ht!]
  \centering
  \includegraphics[width=18cm]{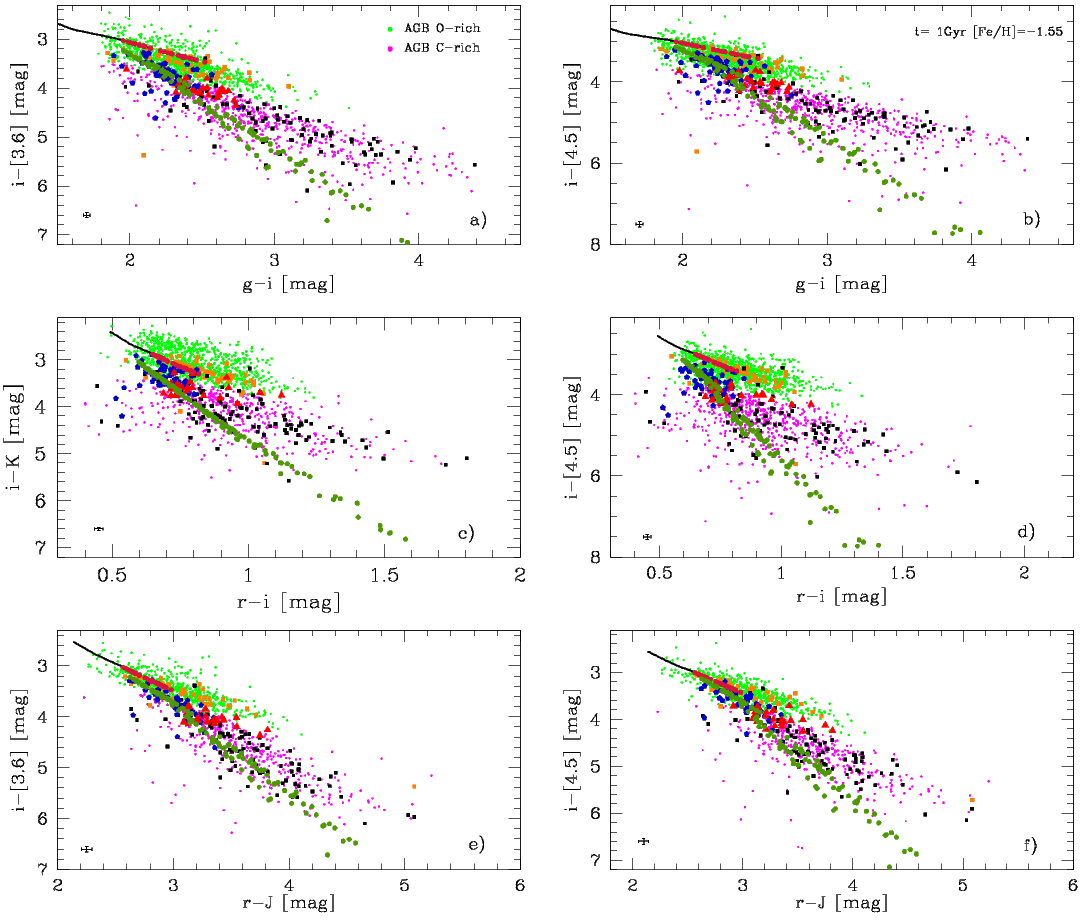}
  \caption{Panel a): Same as the right panels of Figure~\ref{fig:isoage} and Figure~\ref{fig:isomet}, 
           but for the optical-MIR (\textit{i-[3.6]}-\textit{g-i}) CC-D. 
           The PARSEC+COLIBRI stellar isochrone is for an age of 1 Gyr and [Fe/H]=$-1.55$.
           The error bars plotted on the left-bottom side of the CC-D display the intrinsic errors in colors (summed in quadrature).
           Panel b): Same as panel a), but for the optical-MIR (\textit{i-[4.5]}-\textit{g-i}) CC-D.
           The MIR photometry ($[3.6]$ and $[4.5]$ magnitudes) is from \citet{Khan15}.
           Panel c): Same as panel a), but for the optical-NIR (\textit{i-K}-\textit{r-i}) CC-D. 
           Panel d): Same as panel a), but for the optical-MIR (\textit{i-[4.5]}-\textit{r-i}) CC-D. 
           Panel e): Same as panel a), but for the optical-NIR-MIR (\textit{i-[3.6]}-\textit{r-J}) CC-D.
           The NIR photometry ($K$ and $J$ magnitudes) is from \citet{Sibbons12}.
           Panel f): Same as panel a), but for the optical-NIR-MIR (\textit{i-[4.5]}-\textit{r-J}) CC-D.
           \label{fig:AGBcolor}}
\end{figure*}

Figure~\ref{fig:AGBcolor} shows the most indicative CC-Ds defined by 
\citetalias{Tantalo22},  on which we overplotted the
isochrone that properly fit the observed magnitudes and colors 
of TP-AGB stars constructed by assuming t=1 Gyr, [Fe/H]=$-1.55$
and considering the presence of the circumstellar dust.
The isochrones were plotted following the same approach adopted in 
Figures~\ref{fig:isoage}, \ref{fig:isomet}, \ref{fig:ageNodust} and \ref{fig:metNodust}, 
indeed we only display AGB evolutionary phases.  
The newsworthy outcome is that even changing the color combinations among the available optical, NIR and MIR magnitudes, the M- and C-type sequences predicted by the stellar models keep lying on two distinct sequences, 
the separation in magnitude and in color depending on the adopted color combination. 
Theory and observations agree quite well with the photometric sequences 
and the position of the spectroscopic samples on the different CC-Ds. 
This evidence further supports the plausibility of new diagnostics defined in \citetalias{Tantalo22}. 
Moreover, the CC-Ds demonstrate that the \lq{bluer\rq} C-rich stars have the same 
optical-NIR-MIR colors of the observed C-type sample. 

\begin{figure}[ht!]
  \centering
  \includegraphics[width=8.7cm]{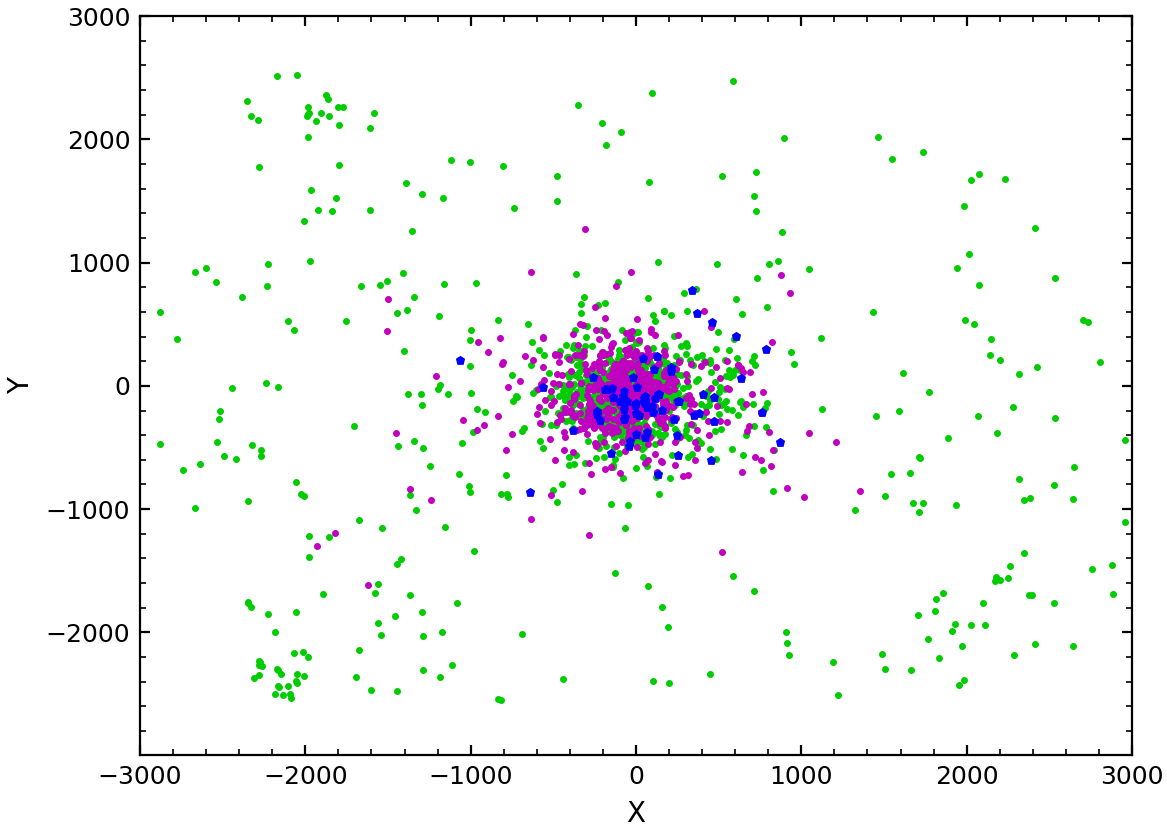}
  \caption{Spatial distribution of the C- (magenta dots) and O-rich (green dots) samples selected from the 
           $(r-K)$-$(r-i)$ CC-D, and the \lq{bluer\rq} C-rich (blue pentagons) selected from the $(Cn-TiO)$-$(R-I)$ CC-D. 
           \label{fig:OCmap}}
\end{figure}

To investigate further the nature of these \lq{bluer\rq} stars we determined their spatial distribution, shown in Figure~\ref{fig:OCmap} together with that of the O- and C-rich samples. As we have widely discussed in \citetalias{Tantalo22} (see Sections~5.3 and 6), O- and C-rich stars have different distributions. The former are extended over the entire FoV, the latter are more centrally concentrated. 
The figure highlights the spatial distribution of the \lq{bluer\rq} C-rich stars being more similar to the C-type group. The evidence from both the CC-Ds and the spatial distribution strongly suggests they can be considered \lq{canonical\rq} C-rich stars.

\begin{figure}[ht!]
  \centering
  \includegraphics[width=8.7cm]{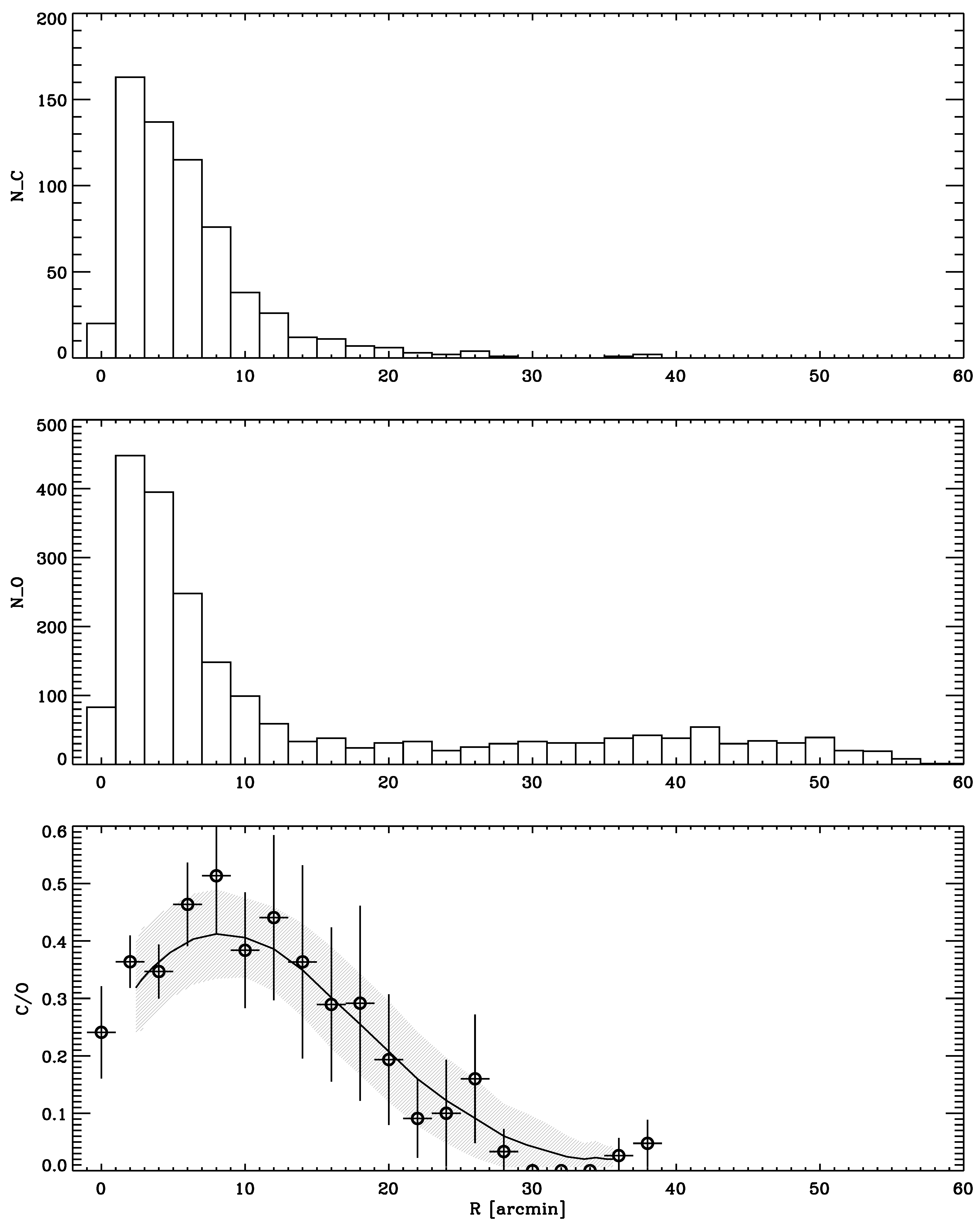}
  \caption{ 
    Top: number of C-rich stars as a function of the radial distance (arcmin). 
    Middle: Same as the top panel, but for O-rich stars. 
    Bottom: Running average of the ration between C- and O-rich stars as a function 
    of the radial distance. See text for more details.
           \label{fig:radial}}
\end{figure}

We have already discussed in \citetalias{Tantalo22} the difference in spatial distribution between C-rich and O-rich stars, but we are now interested in discussing their radial distribution. To estimate the radial distance 
in the galactic restframe we adopted the following formula: 
$$
r_g = \sqrt{x_{ell}^2 + y_{ell}^2}
$$
where $x_{ell}$ and $y_{ell}$ are the elliptical coordinates of both C-rich and O-rich stars, estimated taking into account the position angle and the eccentricity of NGC~6822 (see \citetalias{Tantalo22}).
The top and the middle panels of Figure~\ref{fig:radial} show the radial distribution C- and O-rich stars. As expected the former sample is more centrally concentrated when compared with the latter. The bottom panel of the same figure shows the ratio between C- and O-rich stars. To overcome spurious fluctuations in the ratio due to the adopted radial bin we performed a running average. We first ranked both C- and O-rich stars as a function of the radial distance. We adopted a box of eight arcmin and we estimated the mean radial distance and the ratio of the objects with $r_g$ between zero and eight arcmin. Then we moved outward by two arcmin and estimated the mean radial distance for the objects with $r_g$ between two and ten arcmin. We moved outwards with steps of two arcmin until the farthest object was included in the box. The solid line and the shaded area show the running average and its standard deviation, while the symbols display the ratios of the histograms and their Poissonian uncertainties. The ratio clearly indicates that C-rich stars are vanishing for radial distances larger than 25 arcmin.
In passing, we also note that the decrease in the ratio for radial distances smaller than five arcmin is,
within the errors, quite solid. This circumstantial evidence might indicate the possible presence of a bulge that is mainly dominated by old stellar populations, and in turn, smaller values of the C/O ratio. This working hypothesis requires more detailed investigations in NIR and MIR bands and at high spatial resolution.

In Tables~\ref{tab:Crich} and \ref{tab:Orich} we provide all the optical-NIR-MIR photometric magnitudes and errors of the C- and O-rich samples, respectively. We also provide the ID number of the catalogs, the Gaia DR3 ID number and the sky coordinates.
Note that the C-rich stars come from the selection performed in the 
$(r-K)$-$(r-i)$ CC-D because they make the most populated sample when compared with the selections made on the other CC-Ds (see Section~6 of \citetalias{Tantalo22}). 
The O-rich star sample was obtained from our global optical selection 
of AGB stars ($i$-$(g-i)$ CMD, see Figure~\ref{fig:CMDagb}) once C-rich stars were removed.
A full version of both tables will be supplied in an electronic form.
We only list here the first 5 stars of each catalog for guidance.

Finally, we would like to draw attention on a type of AGB population, called \lq{extreme\rq} AGB stars
(\citealp[or x-AGB; see][and references therein]{Blum06, Srinivasan09, Boyer11, Boyer15b}), dust-enshrouded objects that are experiencing strong mass loss. Accordingly, they are optically obscured, redder, and much more luminous in the IR than  typical AGB stars (M- and C-type). We can roughly identify them in the optical-MIR CMD as a group of stars with colors $(i-[3.6])>6$ mag. Earlier studies asserted that the progenitors of x-AGB stars could have masses around $4-5$ $M_{\odot}$ \citep[i.e.][]{Loon99}. This means that they may be the evolved counterpart of the Cepheids with long periods (P~$\sim$~10d) present in NGC~6822 \citep{Pietr04} and might be associated with populations of ages lower than 500 Myr. However, the latter is only a speculative suggestion and deserves a more quantitative analysis we plan to address in a forthcoming investigation.

\section{Final remarks} \label{sec:conclusion}

In this second paper of our series we have carried out a quantitative photometric analysis of the dIrr galaxy NGC~6822 using the same optical-NIR-MIR dataset presented in \citetalias{Tantalo22}, complemented with deep HST data of two fields in the disk of the galaxy. 
To delve deeper into the nature of the different stellar populations hosted by the galaxy, we compared the observations with a set of isochrones both from the BaSTI-IAC  database and the CMD 3.7 web interface. 
We performed three different selections on our optical catalog based on the spatial distributions of the stellar components provided in \citetalias{Tantalo22}.
We thus obtained three samples representative of the young-, intermediate-, and old-age stellar populations in NGC~6822.
We finally undertook an analysis of the C- and O-rich AGB populations.
Our study has provided the following results:

(i) The comparison between the young sample and the stellar isochrones with different ages 
and metallicities disclosed that this population spans ages from 20 up to 100 Myr and a metal 
content from [Fe/H]=$-0.7$ to [Fe/H]=$-0.4$. These values agree very well with the results 
from previous SFH studies \citep{Gallart96c, Cannon12, Fusco14}.

(ii) The study on the RC stars have revealed that they are very well reproduced by stellar evolutionary models with ages from 4 to 8 Gyr, and metallicities from [Fe/H]=$-1.55$ to [Fe/H]=$-1.3$. This suggests that the intermediate-age stellar populations experienced 
a modest chemical enrichment over an age range of at least 4-8 Gyrs.
We stress that the analysis of the intermediate-age stellar population covers, 
for the first time, the entire body of the galaxy.

(iii) The old population is found to have ages older than 11 Gyr and low metal abundances down to [Fe/H]=$-1.7$. 
Moreover, the CMDs of both the ground-based sample and the HST fields revealed the presence of a highly populated red HB, whose faint edge is matched by a core He-burning sequence for an iron content of [Fe/H]=$-1.55$.
This evidence is once again compatible with the SFH results \citep{Gallart96b, Cannon12}. 
However, we performed a detailed comparison between 
predicted and observed MSTO of the old stellar populations for the very first time.

(iv) The 3D CMD (color $g-r$, magnitude $i$, and the corresponding luminosity function of RG stars) 
and the marginal in $i$-band magnitude of the RG stars allowed us to identify the rarely observed 
AGB clump population, with a luminosity peak of $i\sim 23.35$ mag.
Its location on the $i$-$(g-r)$ CMD ($i\sim 23.35$ mag and $g-r\sim 0.77$ mag) is in good agreement 
with the position of the clump provided by the evolutionary models. Moreover, the latter support 
marginal variations both in magnitude and in color for broad variations in stellar age ($6-13$ Gyr) 
and in  metal content ([Fe/H]=$-1.7/-1.3$). 
It is worth mentioning that the identification of the AGB clump in NGC~6822 is,  
for the first time, fully supported by theoretical predictions.

(v) The analysis of the AGB population has been carried out by comparing the PARSEC+COLIBRI 
isochrones with three diagrams, the optical $i$-($r-i$) and the optical-MIR $[3.6]$-($i-[3.6]$) CMDs, 
plus the optical-MIR ($i-[3.6]$)-($r-i$) CC-D. 
The comparison has revealed that this population spans ages between 1 and 2 Gyr, and  metal content 
between [Fe/H]=$-1.3$ and [Fe/H]=$-1.7$. Looking at the used evolutionary models, we have also 
observed a peculiar behavior, occurring on the optical CMD only, of the isochrones that take account 
of the circumstellar dust during the TP-AGB phase. We found that the latter predict in the optical 
$gri$ bands a slope, for the TP-AGB evolutionary stage, steeper than the observed one, with the 
models becoming increasingly too faint with increasing colors.
This does not happen in the optical-MIR CMD, where models well fit the slope and the whole extension of the AGB. 
Moreover, the same can be seen if we account for isochrones with no dust contribution in both the optical $i$-($r-i$) and optical-MIR $[3.6]$-($i-[3.6]$) CMDs, although they cover a shorter color range.
This evidence suggests that the fairly steep increase in the theoretical magnitudes of the redder AGB stars might be due to too large bolometric corrections applied to the optical filters, when the effect of the circumstellar dust, independently of its composition, is modelled.
The same isochrones, with and without the circumstellar dust effect,
plotted in the optical-MIR ($i-[3.6]$)-($r-i$) CC-D 
clearly show that M- and C-type models 
define two distinct color sequences that agree quite well with both the 
photometric and the spectroscopic samples. The same has been found for the  
other optical-NIR-MIR CC-Ds defined in \citetalias{Tantalo22}. 
The current comparison between theory and observations 
fully validates the newly identified diagnostics to characterize O- and 
C-rich stars in stellar systems.
Moreover, we also demonstrated that the so-called \lq{bluer\rq} C-rich, as 
defined by \citetalias{Letarte02}, 
are truly C-rich AGB stars, since they attain colors in all the optical-NIR-MIR CC-Ds
defined in \citetalias{Tantalo22} and have a spatial distribution similar to the 
\lq{canonical\rq} C-rich stars.
To characterize in more detail the C- and O-rich stars in NGC~6822, we analyzed their
radial distributions finding that, as expected, the former sample is more centrally concentrated
than the latter. We then discussed the variation of the population ratio between C- and O-rich
stars as a function of the radial distance. The results clearly indicate that C-rich stars are
vanishing for radial distances larger than 25 arcmin.
After this validation, the astrometry (Gaia complaint) together 
with optical, NIR and MIR photometry of candidate O- and C-rich stars are made 
available to the astronomical community in electronic form.

Finally, we have speculated that the x-AGB stars with redder colors than the typical  
AGB objects might be the evolved counterpart of Classical Cepheids in NGC~6822, 
associated with stellar populations younger than 500~Myr. This working hypothesis 
is inline with similar suggestions available in literature \citep[i.e.][]{Loon99} 
asserting that their progenitors could have masses of $4-5$ $M_{\odot}$.

\section*{Acknowledgements}

We thank the anonymous referee for his/her positive words concerning the 
content and the cut of the paper, and for his/her pertinent suggestions that 
improved its content and its readability.
It is a real pleasure to thank P. Battinelli and S. Demers for sending
us their catalog of NGC~6822 in electronic form.
M.T. is grateful to Santi Cassisi for his helpful comments and suggestions.
M.T. and M.Monelli acknowledge support from Spanish Ministry of Science, Innovation and Universities (MICIU) through the Spanish State Research Agency under the grant "RR Lyrae stars, a lighthouse to distant galaxies and early galaxy evolution" and the European Regional Development Fun (ERDF) with reference PID2021-127042OB-I00. M. Monelli is also supported by the Severo Ochoa Programe 2020-2023 (CEX2019-000920-S), and by Spanish Ministry of Science, Innovation and Universities (MICIU) through the Spanish State Research Agency under the grant "At the forefront of Galactic Archaeology: evolution of the luminous and dark matter components of the Milky Way and Local Group dwarf galaxies in the Gaia era" with reference PID2020-118778GB-I00/10.13039/501100011033.
M.S. acknowledges support from The Science and Technology Facilities Council Consolidated Grant ST/V00087X/1.
Several of us thank the support from Project PRIN MUR 2022 (code 2022ARWP9C) ‘Early Formation and Evolution of Bulge and HalO (EFEBHO)’ (PI: M. Marconi), funded by the European Union – Next Generation EU, and from the Large grant INAF 2023 MOVIE (PI: M. Marconi). \\
This research has been supported by the Munich Institute for Astro-, Particle and BioPhysics (MIAPbP) which is funded by the Deutsche Forschungsgemeinschaft (DFG, German Research Foundation) under Germany´s Excellence Strategy – EXC-2094 – 390783311. \\
This research has made use of both the BaSTI-IAC database 
(\url{http://basti-iac.oa-abruzzo.inaf.it/index.html}) and the 
CMD 3.7 web interface (\url{http://stev.oapd.inaf.it/cgi-bin/cmd}). \\
It has also made use of the GaiaPortal catalogs access tool, ASI - Space Science Data Center, Rome, Italy (\url{http://gaiaportal.ssdc.asi.it}).\\
This paper is based on data collected at the Subaru Telescope and retrieved from the HSC data archive system, which is operated by the Subaru Telescope and Astronomy Data Center at NAOJ. Data analysis was in part carried out with the cooperation of Center for Computational Astrophysics (CfCA), NAOJ. The HSC collaboration includes the astronomical communities of Japan and Taiwan, and Princeton University. The HSC instrumentation and software were developed by the National Astronomical Observatory of Japan (NAOJ), the Kavli Institute for the Physics and Mathematics of the Universe (Kavli IPMU), the University of Tokyo, the High Energy Accelerator Research Organization (KEK), the Academia Sinica Institute for Astronomy and Astrophysics in Taiwan (ASIAA), and Princeton University. Funding was contributed by the FIRST program from the Japanese Cabinet Office, the Ministry of Education, Culture, Sports, Science and Technology (MEXT), the Japan Society for the Promotion of Science (JSPS), Japan Science and Technology Agency (JST), the Toray Science Foundation, NAOJ, Kavli IPMU, KEK, ASIAA, and Princeton University. \\
All the HST data presented in this paper have been obtained from the Mikulski Archive for Space Telescopes (MAST) at the Space Telescope Science Institute. The specific observations analyzed can be accessed via \dataset[https://doi.org/10.17909/6vkj-nm47]{https://doi.org/10.17909/6vkj-nm47}. STScI is operated by the Association of Universities for Research in Astronomy, Inc., under NASA contract NAS5–26555. Support to MAST for these data is provided by the NASA Office of Space Science via grant NAG5–7584 and by other grants and contracts. \\
This work is based in part on observations made with the Spitzer Space Telescope, which is operated by the Jet Propulsion Laboratory, California Institute of Technology under a contract with the National Aeronautics and Space Administration (NASA). \\
Some of the data reported here were obtained as part of the UKIRT Service Programme. UKIRT is owned by the University of Hawaii (UH) and operated by the UH Institute for Astronomy. When some of the data reported here were obtained, UKIRT was operated by the Joint Astronomy Centre on behalf of the Science and Technology Facilities Council of the U.K. \\

\software{Topcat \citep{Taylor05}, Astropy \citep{Astropy13,Astropy18,Astropy22}, Matplotlib \citep{Hunter07}}

\movetabledown=6.2cm
\begin{rotatetable}
\begin{deluxetable*}{cccccccccccccccccccc}
\tablenum{1}
\tablecaption{Optical-NIR-MIR photometric magnitudes and errors of the current C-rich sample\label{tab:Crich}}
\tablewidth{0pt}
\tabletypesize{\footnotesize}
\tablehead{             
\colhead{ID} & \colhead{Gaia DR3 ID} & \colhead{$\alpha$ (J2000)} & \colhead{$\delta$ (J2000)} & \colhead{$g$} & \colhead{$err_g$} & \colhead{$r$} & \colhead{$err_r$} & \colhead{$i$} & \colhead{$err_i$} & \colhead{$J$} & \colhead{$err_J$} & \colhead{$H$} & \colhead{$err_H$} & \colhead{$K$} & \colhead{$err_K$} & \colhead{[3.6]} & \colhead{$err_{[3.6]}$} & \colhead{[4.5]} & \colhead{$err_{[4.5]}$}
}
\startdata
  278720 & 4182463621278162176 & 296.06425 & -14.73605 & 23.08 & 0.004 & 20.98 & 0.004 & 20.14 & 0.002 & 17.59 & 0.043 & 16.66 & 0.024 & 16.21 & 0.025 & 15.60 & 0.030 & 15.71 & 0.050 \\
  286051 & \nodata		     & 296.07271 & -14.85656 & 22.70 & 0.024 & 20.94 & 0.002 & 20.15 & 0.021 & 17.55 & 0.041 & 16.78 & 0.027 & 16.47 & 0.026 & 16.14 & 0.070 & 16.12 & 0.030 \\
  286422 & 4182451011250681856 & 296.07312 & -14.80169 & 22.45 & 0.005 & 20.71 & 0.011 & 19.97 & 0.007 & 17.62 & 0.043 & 16.74 & 0.027 & 16.46 & 0.029 & 15.86 & 0.050 & 15.97 & 0.050 \\
  288006 & \nodata		     & 296.07487 & -14.84211 & 23.63 & 0.041 & 21.69 & 0.013 & 20.44 & 0.059 & 16.45 & 0.017 & 15.74 & 0.012 & 15.42 & 0.012 & 15.27 & 0.090 & 15.11 & 0.020 \\
  289805 & 4182443967504081536 & 296.07696 & -14.87800 & 22.67 & 0.011 & 20.88 & 0.006 & 19.88 & 0.006 & 17.48 & 0.039 & 16.59 & 0.023 & 16.14 & 0.010 & 15.61 & 0.070 & 15.60 & 0.030 \\
\enddata
\tablecomments{Table~\ref{tab:Crich} is published in its entirety in the machine-readable format. A portion is shown here for guidance regarding its form and content.
The columns from the left to the right are the current ID, Gaia DR3 ID, sky coordinates, optical $gri$ magnitudes with relative errors, NIR $JHK$ magnitudes with relative errors, MIR [3.6] and [4.5] magnitudes with relative errors.}
\end{deluxetable*}
\end{rotatetable}
\movetabledown=6.2cm
\begin{rotatetable}
\begin{deluxetable*}{cccccccccccccccccccc}
\tablenum{2}
\tablecaption{Optical-NIR-MIR photometric magnitudes and errors of the current O-rich sample\label{tab:Orich}}
\tablewidth{0pt}
\tabletypesize{\footnotesize}
\tablehead{             
\colhead{ID} & \colhead{Gaia DR3 ID} & \colhead{$\alpha$ (J2000)} & \colhead{$\delta$ (J2000)} & \colhead{$g$} & \colhead{$err_g$} & \colhead{$r$} & \colhead{$err_r$} & \colhead{$i$} & \colhead{$err_i$} & \colhead{$J$} & \colhead{$err_J$} & \colhead{$H$} & \colhead{$err_H$} & \colhead{$K$} & \colhead{$err_K$} & \colhead{[3.6]} & \colhead{$err_{[3.6]}$} & \colhead{[4.5]} & \colhead{$err_{[4.5]}$}
}
\startdata
  355961 & \nodata		     & 296.13787 & -14.76861 & 22.64 & 0.005 & 21.12 & 0.006 & 20.36 & 0.002 & 18.31 & 0.068 & 17.45 & 0.048 & 17.24 & 0.050 & 17.09 & 0.050 & 17.21 & 0.050 \\
  356584 & 4182465438045963392 & 296.13833 & -14.66700 & 22.12 & 0.018 & 20.37 & 0.002 & 19.32 & 0.011 & 16.88 & 0.020 & 15.92 & 0.010 & 15.71 & 0.013 & 15.50 & 0.070 & 15.64 & 0.030 \\
  357611 & 4182446888084533888 & 296.13908 & -14.85591 & 22.57 & 0.004 & 21.09 & 0.003 & 20.02 & 0.003 & 17.72 & 0.043 & 17.12 & 0.037 & 16.96 & 0.041 & 16.98 & 0.210 & 16.45 & 0.030 \\
  358159 & 4182439530802291200 & 296.13950 & -14.94908 & 21.52 & 0.011 & 19.96 & 0.001 & 19.31 & 0.007 & 17.14 & 0.027 & 16.47 & 0.020 & 16.26 & 0.025 & 16.06 & 0.050 & 16.20 & 0.050 \\
  358437 & \nodata		     & 296.13979 & -14.72897 & 22.60 & 0.003 & 20.92 & 0.004 & 20.12 & 0.003 & 17.80 & 0.045 & 16.99 & 0.034 & 16.72 & 0.032 & 16.56 & 0.070 & 16.63 & 0.070 \\
\enddata
\tablecomments{Table~\ref{tab:Orich} is published in its entirety in the machine-readable format. A portion is shown here for guidance regarding its form and content.
The columns from the left to the right are the current ID, Gaia DR3 ID, sky coordinates, optical $gri$ magnitudes with relative errors, NIR $JHK$ magnitudes with relative errors, MIR [3.6] and [4.5] magnitudes with relative errors.}
\end{deluxetable*}
\end{rotatetable}

\clearpage

\appendix

\section{Precision of the current photometry} \label{appendix}

There are different softwares available in the astronomical community to perform 
accurate PSF photometry in crowded stellar fields.
Among the different options there are DoPHOT by \citet{Schechter93} and an updated version by \citet{Alonso12}, and DOLPHOT by \citet{Dolphin00,Dolphin16}.
The reason why we are using DAOPHOT/ALLFRAME is threefold.
a) We can extract all the information available in the images, since the position of 
the stellar centroids is performed simultaneously over all the multi-band images 
available. This typically means an improved accuracy in crowded stellar fields 
when approaching the limiting magnitude.
b) The individual images are calibrated using local standards. This typically means 
a better accuracy in the photometric zero-points when compared with the calibration 
of the mean magnitude. 
c) This approach provide the unique opportunity to perform a detailed search 
for variable stars, since the extracted time series are homogeneous not only 
for the multi-band photometry, but also for the absolute calibration.
The reader interested in a more detailed discussion concerning aperture and 
PSF photometry is referred to \url{https://en.wikipedia.org/wiki/Photometry_(astronomy)}.

We now discuss the precision and the uniformity of the photometric zero-points over the FoV.
In the final photometric combination of all the data for NGC~6822, the individual
photometric zero point of each image is redetermined from a local network of
3,264 local standard stars within our NGC~6822 target field.  The magnitudes and
colors of these stars have been established by direct comparison to an all-sky
network of photometric standards under development by P.B.~Stetson \citep[e.g.][]{Stetson00,Stetson05,Stetson19} 
over approximately the last 30 years, as first mentioned in \citet*{Stetson98}.  
Most of this work is based upon other
people's public-domain data downloaded from the various on-line astronomical
archives, but a significant fraction is derived from targeted observations
obtained by us for this purpose.

As of the moment of this writing (2024 December 11), the database from which
this photometric standard network is derived consists of 11,599 individual
datasets, where a ''dataset`` is defined as the output of one CCD on one night;
an eight-chip mosaic, like WFI on the ESO/MPI 2.2m telescope for instance, would
produce eight datasets per night. These datasets comprise somewhat over
800,000 individual CCD images from some 4,101 nights distributed among 880
observing runs over 41.6 years (1983 January 8 through 2024 August 19).  These
data were obtained with (at the moment) 49 different telescopes and a greater
number of different instruments.

Each dataset is individually calibrated to our current approximation of the
Landolt \citep{Landolt73,Landolt92} realization of the Johnson {\it UBV\/} 
and Kron-Cousins {\it RI\/} photometric systems.  The transformation
equation for each dataset normally includes linear and quadratic color terms,
and for larger-format CCDs linear and often quadratic terms in the spatial
location of the star on the detector.  Normally, mean values for the color terms
are determined and employed for all the datasets of a given observing run---sometimes 
collectively for the individual chips in mosaic cameras, sometimes separately
depending upon the transformation residuals---but zero points and spatial
transformation terms are always derived for each individual dataset.  On nights
where photometric conditions apparently prevailed a nightly global zero point and
atmospheric extinction coefficients are derived; the extinction coefficients are
derived collectively and applied uniformly to all the CCDs in a multi-chip mosaic,
but each chip gets its own individual nightly zero point.  On apparently
non-photometric nights where nevertheless useful observations had been obtained,
no global zero point or extinction coefficients were derived.  Instead, a
specific zero point is computed each for each individual CCD image containing
more than one standard.  These non-photometric data do not improve the absolute
zero-point of all stars contained in a particular CCD-sized patch of sky, but
they do help to reduce random photometric errors relative to each other among
that particular subset of stars.  Of the 11,599 datasets reduced at this moment,
we consider that 5,318 were obtained under photometric conditions, and 6,281
have been reduced in non-photometric mode.

A comparison of Landolt's published photometry with our own average photometry
of his stars, included within our database, show root-mean-square (rms) differences per
star of 0.0143 mag in $B$ (307 stars in common), 0.0102 mag in $V$ (320 stars),
0.0102 mag in $R$ (219 stars), and 0.0137 mag in $I$ (221 stars).  These values
appear to represent the ultimate impossibility of duplicating the detailed
wavelength sensitivity of typically available filter/detector bandpasses, as they
measure the wide variety of possible stellar spectral-energy distributions.  We
feel, however, that overall our net standard-star network should be on Landolt's
zero points at a level of order 0.001 mag in each filter.  Nightly
rms residuals of observed magnitudes of individual stars from {\it our\/} 
adopted standard values are typically $<$ 0.02 mag in each filter. 
Specifically, the median values of the rms star-to-star calibration residual
over all available datasets are:  0.0121 mag in $B$ based upon 8,306 datasets
containing $B$ observations, 0.0094 mag in {\it V\/} (9,412 datasets), 0.0081
mag in $R$ (6,418 datasets), and 0.0110 mag in $I$ (8,252 datasets);
inter-quartile ranges are 0.0076--0.0182 mag in $B$, 0.0059--0.0139 mag in $V$,
0.0045--0.0135 mag in $R$, and 0.0062--0.0171 mag in $I$.  These numbers
probably reflect (a)~random vagaries of the instantaneous atmospheric extinction
from the simple relation in different directions at different times, and
(b)~inadequate flat-fielding of the observed images, as well as the
aforementioned (c)~idiosyncrasies of individual filter/detector bandpasses as
they sample different spectral-energy distributions.  As context, among the
data considered here, then median number of standard-star observations used to
calibrate an individual dataset (i.e., the median number of standard-star
observations per filter per CCD per night) is 1,881 in $B$, 3,412 in $V$, 2,684
in $R$, and 2,567 in $I$.  Star-to-star residuals of order 0.02 mag divided by
the square root of numbers of order $10^3$ suggests that the uncertainty of a
typical calibration is small compared to the random uncertainty of any
individual stellar measurement. Our photometry is available to concerned
researchers who want a more detailed comparison of our photometric system with Landolt's more recent publications or with other photometric studies of our target fields.

At this precise moment the all-sky standard star network includes some 273,102 stars
meeting the following {\it minimal\/} standards: at least five observations in a
given filter on photometric occasions, a standard error of the mean magnitude
in a given filter no greater than 0.02 mag, and no evidence of intrinsic
variability in excess of 0.05 mag, rms, considering all available
filters together.  However, this standard-star network evolves incrementally,
mostly as new observing runs are added, but also to a lesser extent as
calibration software and methodology improve, as new stars become acceptable as
standards due to new data, and as apparently variable stars are recognized and
removed from the list of accepted standard stars.  The instantaneous state of
this standard-star system in (at the moment) 364 target fields on the sky can be
freely obtained by anyone at any time\footnote{They are available at \url{https://www.canfar.net/storage/list/STETSON/homogeneous/Latest_photometry_for_targets_with_at_least_BVI}
and \url{https://www.canfar.net/storage/vault/list/STETSON/homogeneous/Latest_photometry_for_targets_with_only_V_and_B_or_R_or_I}}.

Interested readers are welcome to download whatever they can use.  We encourage
them to retain older copies of the data files as new ones are made available, so
that they can monitor the evolution of the network and judge whether that
affects their science.  We find that the adopted photometric magnitudes of a
reasonably well-observed star typically change only in the third or fourth decimal
place between subsequent generations of analysis.

As mentioned above, in the last step before combining the accumulated measured
magnitudes into final photometric indices for target stars, the individual
photometric zero point assigned to each CCD image is reconsidered on the basis
of local photometric standards belonging to the larger all-sky network that happen
to be contained within that image.  In the particular case of NGC~6822,
this means that zero points have been redetermined for 9,124 images:  2,884 $B$,
2,087 $V$, 1,085 $R$, and 3,068 $I$.  Among these, the median number of local
standards used to determine the personal zero point of an individual CCD image was 154
(interquartile range 70--300); the median standard error of a mean zero point
obtained in this way was 0.0009 mag, interquartile range 0.0007--0.0017.
More specifically, the median uncertainty of a mean zero point derived for
an individual image was 0.0010 mag in $B$, 0.0007 mag in $V$, 0.0013 mag in $R$,
and 0.0010 mag in $I$.  The accumulated error due to uncertain zero points
for a star measured in more than a few different images should therefore be small.
To be sure, this uncertainty will not be the same at every point within the target area,
but we do not expect this component of the error budget to vary strongly around the field:
the center of the area appears in more different images than the periphery, but on the
other hand measurements in the center are more compromised by crowding than those around the
edge.  These two effects should partially compensate each other.  Overall, within our
studied area 90\% of stars brighter than $V$ = 23.5 appear in at least 35 $B$ images (median
96, maximum 395), 9 $V$ images (54, 267), 18 $R$ images (51, 131), and 25 $I$ images
(82, 314).  As we consider brighter magnitude limits, the 90-th and 50-th percentile numbers
increase.  For $V$ < 22.5, for instance, 90\% of stars appear in at least 17 $V$ images, and
half appear in at least 88.

Readers in need of further details concerning our analysis are invited to communicate with co-author P.B.~Stetson.

\bibliographystyle{aasjournal}



\end{document}